\documentclass[article,traditabstract]{aa}
\usepackage{txfonts}
\usepackage{color}
\usepackage{graphicx}
\usepackage{natbib}
\usepackage{multirow}
\usepackage[colorlinks=true, allcolors=blue]{hyperref}

\begin{document}

  \title{The Ring Structure in the MWC 480 Disk Revealed by ALMA}
  \author{Yao Liu$^{1,2}$
           \and
           Giovanni Dipierro$^{3}$ 
           \and 
           Enrico Ragusa$^{4}$
           \and
           Giuseppe Lodato$^{4}$
           \and
           Gregory J. Herczeg$^{5}$
           \and
           Feng Long$^{5}$
           \and
           Daniel Harsono$^{6}$
           \and
           Yann Boehler$^{7,8}$
           \and
           Francois Menard$^{8}$
           \and
           Doug Johnstone$^{9,10}$
           \and
           Ilaria Pascucci$^{11,12}$
           \and
           Paola Pinilla$^{13}$
           \and
           Colette Salyk$^{14}$
           \and
           Gerrit van der Plas$^{8}$
           \and
           Sylvie Cabrit$^{8,15}$
           \and
           William J. Fischer$^{16}$
           \and
           Nathan Hendler$^{11}$
           \and
           Carlo F. Manara$^{17}$
           \and
           Brunella Nisini$^{18}$
           \and
           Elisabetta Rigliaco$^{19}$
           \and
           Henning Avenhaus$^{1}$
           \and
           Andrea Banzatti$^{11}$
           \and
           Michael Gully-Santiago$^{20}$}
   \institute{Max Planck Institute for Astronomy, K\"onigstuhl 17, D-69117 Heidelberg, Germany; \email{yliu@mpia.de}
            \and
            Purple Mountain Observatory \& Key Laboratory for Radio Astronomy, Chinese Academy of Sciences, 2 West Beijing Road, Nanjing 210008, China
            \and
            Department of Physics and Astronomy, University of Leicester, Leicester LE1 7RH, United Kingdom
            \and
            Dipartimento di Fisica, Universit$\grave{\rm a}$ Degli Studi di Milano, Via Celoria, 16, Milano, I-20133, Italy
            \and
            Kavli Institute for Astronomy and Astrophysics, Peking University, Yiheyuan 5, Haidian Qu, 100871 Beijing, China
            \and
            Leiden Observatory, Leiden University, P.O. box 9513, 2300 RA Leiden, The Netherlands
            \and
            Rice University, Department of Physics and Astronomy, Main Street, 77005 Houston, USA 
            \and
            Univ. Grenoble Alpes, CNRS, IPAG, F-38000 Grenoble, France
            \and
            NRC Herzberg Astronomy and Astrophysics, 5071 West Saanich Road, Victoria, BC, V9E 2E7, Canada
            \and
            Department of Physics and Astronomy, University of Victoria, Victoria, BC, V8P 5C2, Canada
            \and 
            Lunar and Planetary Laboratory, University of Arizona, Tucson, AZ 85721, USA
            \and 
            Earths in Other Solar Systems Team, NASA Nexus for Exoplanet System Science, USA
            \and
            Department of Astronomy/Steward Observatory, The University of Arizona, 933 North Cherry Avenue, Tucson, AZ 85721, USA
            \and
            Vassar College Physics and Astronomy Department, 124 Raymond Avenue, Poughkeepsie, NY 12604, USA
            \and
            Sorbonne Universit\'{e}, Observatoire de Paris, Universit\'{e} PSL, CNRS, LERMA, F-75014 Paris, France
            \and
            Space Telescope Science Institute Baltimore, MD 21218, USA
            \and
            European Southern Observatory, Karl-Schwarzschild-Str. 2, D-85748 Garching bei M\"{u}nchen, Germany
            \and
            INAF-Osservatorio Astronomico di Roma, via di Frascati 33, 00040 Monte Porzio Catone, Italy
            \and
            INAF-Osservatorio Astronomico di Padova, Vicolo dell'Osservatorio 5, 35122 Padova, Italy 
            \and
            NASA Ames Research Center and Bay Area Environmental Research Institute, Moffett Field, CA 94035, USA}
\authorrunning{Liu et al.}
\titlerunning{The Ring Structure in the MWC 480 Disk}
  
\abstract{Gap-like structures in protoplanetary disks are likely related to planet formation processes. In this paper, we present and analyze high 
resolution ($0.17^{\prime\prime}\times 0.11^{\prime\prime}$) 1.3\,mm ALMA continuum observations of the protoplanetary disk around the Herbig Ae 
star MWC\,480. Our observations for the first time show a gap centered at ${\sim}\,74\,\rm{au}$ with a width of ${\sim}\,23\,\rm{au}$, surrounded 
by a bright ring centered at ${\sim}\,98\,\rm{au}$ from the central star. Detailed radiative transfer modeling of both the ALMA image and the broadband 
spectral energy distribution is used to constrain the surface density profile and structural parameters of the disk. If the width of the gap corresponds 
to $4\,{\sim}\,8$ times the Hill radius of a single forming planet, then the putative planet would have a mass of $0.4\,{\sim}\,3\,\rm{M_J}$. We 
test this prediction by performing global three-dimensional smoothed particle hydrodynamic gas/dust simulations of disks hosting a migrating and 
accreting planet. We find that the dust emission across the disk is consistent with the presence of an embedded planet with a mass 
of ${\sim}\,2.3\, \mathrm{M_{\mathrm{J}}}$ at an orbital radius of ${\sim}\,78\,\rm{au}$. Given the surface density of the best-fit radiative 
transfer model, the amount of depleted mass in the gap is higher than the mass of the putative planet, which satisfies the basic condition 
for the formation of such a planet.}

\keywords{protoplanetary disks -- planet-disk interactions -- radiative transfer -- stars: formation -- stars: individual (MWC 480)}

\maketitle

\section{Introduction}
\label{sec:intro}

Protoplanetary disks are a natural outcome of the angular momentum conservation during the collapse of a rotating molecular cloud \citep{terebey1984,shu1987}. 
These disks can be approximately characterized by a hot surface layer and a cool interior region close to the disk midplane \citep[e.g.,][]{calvet1991,chiang1997,dullemond2001}. 
Imaging disks at different wavelengths provides insights into a large number of fundamental physical properties of the disk population, which allow us to put 
constraints on models of disk evolution and planet formation. Observations in the infrared (IR) spectral range probe the distribution of small dust grains 
in the disk surface layer, whereas disk emission in the (sub-)millimeter regime is dominated by the thermal emission from large dust grains populating the 
regions close to the midplane, as a combined consequence of grain growth and vertical settling \citep[e.g.,][]{miyake1995,dalessio2001,dullemond2004,birnstiel2016,pinte2016,louvet2018}. 
Investigating the detailed structure of the disk, in particular the disk interior and dense midplane probed by high-resolution millimeter observations, plays a 
key role in building the picture of planet formation and disk evolution. 

In recent years, with its unprecedented sensitivity and spatial resolution, the Atacama Large Millimeter/submillimeter Array (ALMA) has revealed a 
series of interesting structures in protoplanerary disks, such as gaps and rings \citep[e.g.,][]{alma2015,andrews2016,cieza2017, dipierro2018,fedele2018}, 
horseshoe structures \citep[e.g.,][]{casassus2013,vanderMarel2013,kraus2017}, and spiral arms \citep{tobin16a,perez2016}. With the increasing number of targets 
observed by ALMA, gaps and rings appear to be prevalent in disks around both T Tauri and Herbig stars regardless of their age \citep[e.g.,][]{zhang16a,isella2016, hendler2017, vanderplas2017}. 
Several mechanisms have been proposed to explain the origin of the gap/ring structure, including sintering-induced dust rings \citep{okuzumi2016}, dust coagulation/fragmentation 
triggered by condensation zones of volatiles \citep{zhang2015,banzatti2015,pinilla2017}, dead zones \citep{pinilla2016}, the operation of a secular gravitational 
instability \citep{takahashi2016}, planet-disk interactions \citep[e.g.,][]{dipierro2015,jin2016,rosotti2016,dipierro2018,fedele2018}, and 
non-ideal magnetohydrodynamics (MHD) effects \citep{bethune2016}.

In this paper, we present high angular resolution ALMA images of 1.3\,mm continuum emission from the protoplanetary disk around MWC\,480 (also frequently called HD 31648), 
and subsequently model the disk with hydrodynamic simulations to infer the mass of a possible planet within the disk. %The 1.3\,mm image of MWC 480 shows dark and bright 
%rings (or a gap and emission ring) at $\sim$74 and $\sim$98\,au from the central star, which have not been identified in previous millimeter observations.
MWC\,480 is a Herbig Ae star located at a Gaia DR2 distance\footnote{This distance is consistent with the distance of $160\,\rm{pc}$, calculated from the weighted mean 
of Gaia DR2 parallax measurements, of the 16 previously-identified Taurus members \citep{luhman2017} that are located within one degree of MWC\,480. All values from the 
literature used in this paper are updated to this new distance, from previous distance estimates of ${\sim}\,140$\,pc.} of $161.8_{-2}^{+2}\,\rm{pc}$ \citep{gaia2016,gaia2018} 
with an age of ${\sim}\,7$\,Myr \citep{simon2000} in the Taurus star forming region. The mid-to-far IR spectral energy distribution (SED) is consistent with a Group II Meeus 
disk \citep{juhasz2010}, following the grouping schemes of \citet{meeus2001} and \citet{vanboekel2005}. The SED does not show any evidence for an inner hole or an opacity gap, 
as there is no sharp flux deficit in the near-IR and/or mid-IR \citep{espaillat2007,espaillat2010,williams2011}. Previous (sub-)millimeter observations suggested the presence 
of a bright emission from a smooth disk without significant substructures \citep{hamidouche2006,pietu2006,guilloteau2011,huang2017}. Polarization images of scattered light 
off the disk in H-band reveals that the small grains extend to ${\sim}\,160$\,au and are smoothly flared, with no significant substructures detected with a spatial resolution 
of $0.07^{\prime\prime}$ \citep{kusakabe2012}. The Keplerian rotation measured in bright CO rotational line emission from the disk leads to a mass 
of ${\sim}\,2.0$\,M$_{\odot}$ \citep{simon2000,pietu2007}, after correcting for the updated distance. The star is actively accreting from 
the disk at a variable rate of ${\sim}\,1.2\times10^{-7}$ M$_\odot$/yr (\citealt{mendigutia2012}; see also, e.g., \citealt{donehew2011} and \citealt{salyk2013}).

The MWC\,480 disk is selected for this in-depth study as an initial result from an ALMA survey of 32 disks in the Taurus Molecular Cloud because of its prominent ring 
and gap structures. The observational details are given in Sect.~\ref{sec:obs}. Sect.~\ref{sec:rtmodeling} describes our radiative transfer modeling with the goal of 
constraining the surface density profile and structural parameters of the disk. In Sect.~\ref{sec:hydromodeling}, hydrodynamics simulations are performed to explore 
planet-disk interaction as the origin of the gap. We briefly discuss other potential interpretations of the ring structure in Sect.~\ref{sec:discussion}, followed by 
a summary in Sect.~\ref{sec:summary}.

\begin{figure*}[!t]
 \centering
 \includegraphics[width=0.98\textwidth]{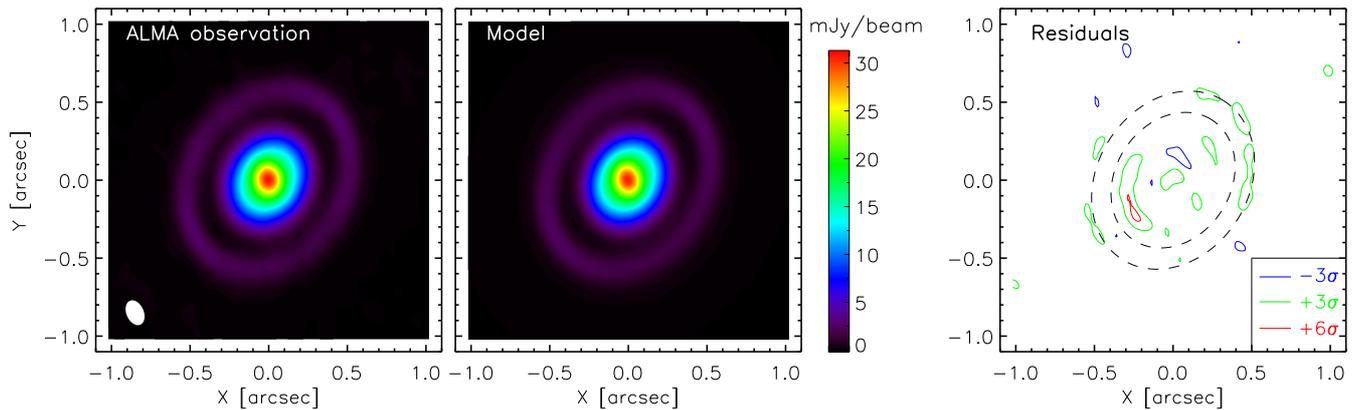}
 \caption{Comparison between the ALMA observation (left) and the model image (middle) at Band 6. The beam size is indicated by the white filled ellipse in the observed image. 
The residuals between the observed and model maps are shown in the right panel, in which $\sigma$ refers to the rms level $0.07\,\rm{mJy/beam}$ of the CLEANed map. The dashed 
ellipses indicate the locations of the gap and ring.} 
 \label{fig:obsimg}
\end{figure*}

\begin{figure*}[!t]
  \centering
  \includegraphics[width=0.80\textwidth]{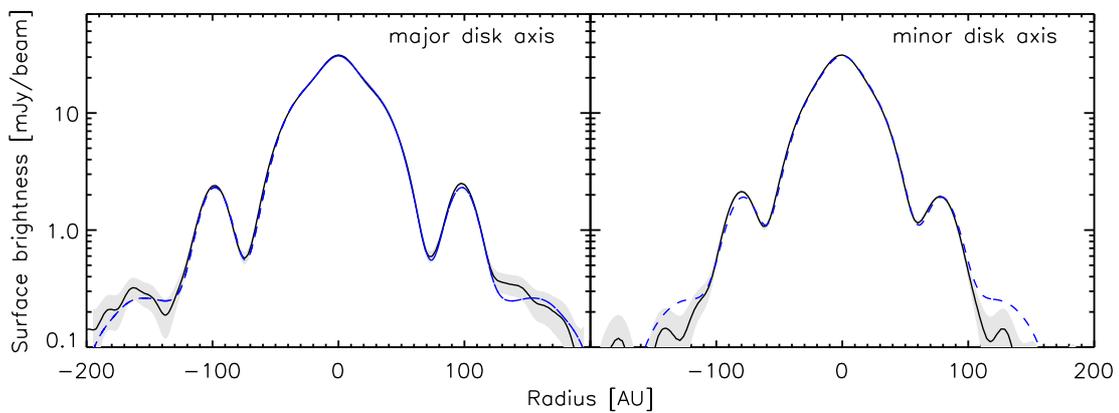}
  \caption{Comparison between the observed brightness distribution and model prediction along the disk major (left) and minor (right) axis. Black solid lines show the 
observation with uncertainties indicated as gray shaded region, whereas blue dashed lines refer to model predictions.}
  \label{fig:obsflux}
\end{figure*}

\section{Observation, data reduction and analysis}
\label{sec:obs}

We observed MWC\,480 with ALMA in Band 6 (230\,GHz) as part of the Cycle 4 program (ID: 2016.1.01164.S; PI: G. Herczeg) on August 27, 2017. 
The observation was conducted using 47 12\,m antennas with baselines ranging from 21 to 3638\,m. The on-source time was 4 minutes and  
the precipitable water vapor (PWV) was 0.5\,mm during the integration. J0510+1800 served as the flux calibrator, while J0510+1800 and 
J0512+2927 were observed for bandpass and phase calibration.

The data were calibrated using the Common Astronomy Software Applications (CASA, \citealt{mcmullin07a}) package, version 5.1.1. Following the data reduction scripts provided 
by ALMA, the atmospheric phase noise was first reduced using water vapor radiometer measurements. Then the standard bandpass, flux, and gain calibrations were applied accordingly. 
Based on the phase and amplitude variations on calibrators, we estimated an absolute flux calibration uncertainty of ${\sim}\,10$\%. Self-calibration was performed to improve 
the signal-to-noise ratio. The continuum image was finally created from the calibrated visibilities using the \texttt{CASA} \texttt{tclean} task with Briggs weighting with a 
robust parameter of 0.5. The resulting beam size is $0.17^{\prime\prime}\times0.11^{\prime\prime}$, corresponding to a physical scale of $27.5 \times 17.8$ au at a distance 
of 161.8\,pc. The rms noise level of the image is ${\sim}\,0.07\,\rm{mJy/beam}$. %{\bf The signal-to-noise ratio of the peak beam brightness is as high as ${\sim}$~450.} 

The cleaned image of 1.3\,mm continuum emission (left panel of Figure~\ref{fig:obsimg}) reveals a pair of gap and ring.  Figure~\ref{fig:obsflux} displays the radial brightness 
distribution along the major and minor axis of the disk, assuming a position angle (PA) of $148^{\circ}$ obtained from the \texttt{CASA} \texttt{imfit} task. Modeling the visibility 
data yields a similar PA value and a total flux density at 1.3\,mm of $268\,{\pm}\,0.3\,\rm{mJy}$ \citep{long2018}. The radial extent of the disk is 
${\sim}\,175$\,au, defined as the radius at $3\,{\times}\,{\rm rms}$ noise level of the image (i.e., 0.21\,mJy/beam). The observed gap is deeper along the major axis than the minor axis. 
This difference in gap depth might be introduced artificially because the asymmetric beam ($\rm{PA}\,{\sim}\,23^{\circ}$) has a major axis that is closely aligned with the minor axis 
of the disk, or may be the consequence of the projection effect of a geometrically-thick disk.

Fitting a Gaussian profile to the brightness distribution along the major axis in the vicinity of the gap using the \texttt{mpfit} routine within \texttt{IDL} yields a gap 
location of $74.3\,\rm{au}$ ($0.46^{\prime\prime}$) and a full-width half-maximum (FWHM) of $23\pm3\,\rm{au}$. The same procedure yields a center for the bright ring 
of $97.5\,\rm{au}$ ($0.6^{\prime\prime}$). There is a small hint of a second ring at ${\sim}\,1^{\prime\prime}$, but it has a low significance. Follow-up observations
are required to confirm the existence and place constraints on its properties.

\section{Radiative transfer modeling}
\label{sec:rtmodeling}

In order to constrain the surface density profile and structural parameters of the MWC 480 disk, we perform radiative transfer modeling of the 
broadband SED and the ALMA image using the radiative transfer code \texttt{RADMC-3D}\footnote{http://www.ita.uni-heidelberg.de/~dullemond/software/radmc-3d/.} \citep{radmc3d2012}. 
The disk is considered to be passively heated by stellar irradiation. The construction of the surface density is linked to the brightness distribution (see Sect.~\ref{sec:surdens}),
because most parts of the disk are optically thin at 1.3\,mm (see Figure~\ref{fig:temp2d}). Therefore, our modeling of the ALMA data is carried out in the image plane. 
However, we confirm the quality of the fit in the UV space. Table~\ref{tab:parameter} summarizes the model parameters.

We employ a two-dimensional flared disk model, which has been successfully used to explain observations of protoplanetary disks \citep[e.g.,][]{wolfp2003,sauter2009,lium2012,liu2017}. 
The disk model includes distinct populations of small grains, used to set up the temperature structure and to reproduce the IR SED, and of large grains to reproduce the 1.3\,mm dust 
continuum image. The density structure of both grain populations is fully parameterized rather than solved under an assumption of hydrostatic equilibrium. In particular, the 
flaring index and scale height are free parameters that are not self-consistently derived from the temperature calculation (see Sect.~\ref{sec:dustdens} and Sect.~\ref{sec:surdens}). 
This approach has an advantage that one can directly assess the effect of each model parameter on the simulated observable.
The dust in the model is a mixture of 75\% amorphous silicate and 25\% carbon. For the complex refractive indices, we take the data from \citet{dorschner1995} for 
silicate and from \citet{jager1998} for carbon at a pyrolysis temperature of $800\,^{\rm o}\rm{C}$. Mie theory is used to calculate the dust opacities. The grain size distribution 
follows the standard power law ${\rm d}n(a)\propto{a^{-3.5}} {\rm d}a$ with a minimum grain size of $a_{\rm{min}}=0.01\,\mu{\rm m}$. The maximum grain size is set 
to $a_{\rm{max}}\,{=}\,2\,\mu\rm{m}$ for the small grain population (SGP) and $a_{\rm{max}}\,{=}\,3\,\rm{mm}$ for the large grain population (LGP). The mass fraction of large 
dust grains relative to the total dust mass in the disk ($M_{\rm dust,tot}$) is described as $f_{\rm LGP}$, so that small grains have a mass fraction of $1-f_{\rm LGP}$.

\begin{figure}[!t]
\includegraphics[width=0.48\textwidth]{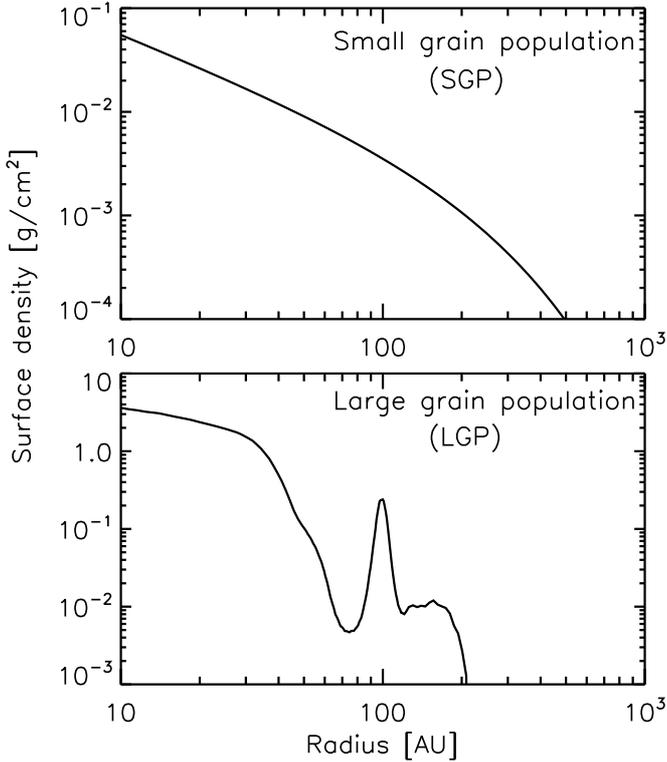}
\caption{Surface density profiles of the best-fit radiative transfer model. {\it Upper: } The surface density of the small grain population. It follows the profile described 
in Eq.~\ref{eqn:sigma} with $R_{\rm{C}}=200\,\rm{au}$ and $\gamma=1$. {\it Bottom: } The surface density of the large grain population. It is obtained by 
fitting the brightness distribution along the disk major axis, see Sect.~\ref{sec:surdens}.}
\label{fig:surdens}
\end{figure}

\subsection{The distribution of small grains}
\label{sec:dustdens}

The surface density profile of the small grain population, $\Sigma_{\rm SGP}$, is modeled as a power law with an exponential taper, as 
\begin{equation}
\Sigma_{\rm SGP}(R) \propto \left(\frac{R}{R_{\rm C}}\right)^{-\gamma}{\rm exp}\left[-\left(\frac{R}{R_{\rm C}}\right)^{2-\gamma}\right],
\label{eqn:sigma}
\end{equation}
where $R_{\rm C}\,{=}\,200\,\rm{au}$ is a characteristic radius and $\gamma\,{=}\,1$ is the gradient parameter. The vertical structure of the small grains 
is modeled as a Gaussian profile
\begin{equation}
\rho_{\rm{SGP}} (R,z)\propto\frac{\Sigma_{\rm SGP}(R)}{h_{\rm SGP}(R)}\,\exp\left[-\frac{1}{2}\left(\frac{z}{h_{\rm SGP}(R)}\right)^2\right], \\
\label{eqn:dens}
\end{equation}
where $R$ is the radial distance from the central star measured in the disk midplane. 
We fix the disk inner radius to $R_{\rm in}\,{=}\,0.15\,\rm{au}$, which is consistent with the result from models of the SED \citep{fernandes2018} and  
the near- and mid-IR interferometric observations \citep{millan-gabet2016}. The outer radius of the small grain population is set to $750\,\rm{au}$, close 
to the value from $^{12}{\rm CO}$ observations \citep{pietu2007}, since small grains are expected to be well coupled with the gas. 

The scale height of the small grain population, $h_{\rm SGP}$, follows a power-law profile
\begin{equation}
h_{\rm SGP} = H_{100}\times\left(\frac{R}{100\,\rm{au}}\right)^\beta,\\
\label{eq:heightgas}
\end{equation}
with $\beta$ characterizing the degree of flaring and $H_{100}$ representing the scale height at a distance of $100\,\rm{au}$.

\begin{table}[!t]
\caption{Model parameters.}
\centering
\linespread{1.3}\selectfont
\begin{tabular}{lcc}
\hline 
\texttt{Stellar parameters} & Value &  Fixed/free \\
\hline
$M_{\star}\, [M_{\odot}]$  & 2.0         &  fixed \\
$T_{\rm eff}\, [\rm K]$ $^a$   & 8460    &  fixed \\
$L_{\star}\, [L_{\odot}]$ $^b$ & 17.3    &  fixed \\
$A_{V}$\,[mag] $^b$            & 0.1     &  fixed \\
\hline
\texttt{Disk parameters} & & \\
\hline
$R_{\rm in}\,[\rm{au}]$                &  0.15                    &    fixed         \\
$R_{\rm out.SGP}\,[\rm{au}]$           &  750                     &    fixed         \\
$R_{\rm out.LGP}\,[\rm{au}]$           &  200                     &    fixed         \\
$\gamma$                               &  1.0                     &    fixed         \\
$\beta$                                &  $1.08^{+0.02}_{-0.02}$        &    free          \\
$H_{100}\,[\rm{au}]$                   &  $12^{+2.1}_{-0.7}$            &    free          \\ 
$M_{\rm dust,tot}\, [10^{-3}\,M_{\odot}]$  &  $1.6^{+0.5}_{-0.4}$       &    free          \\
$\Lambda$                              &  $0.25^{+0.04}_{-0.04}$        &    free          \\
$f_{\rm LGP}$                          &  $0.95^{+0.04}_{-0.08}$        &    free          \\
\hline 
\texttt{Observational parameters} & & \\
\hline
$i\,[^{\circ}]$                &  $37^{+0.3}_{-0.8}$     &   free    \\
Position angle\,[$^{\circ}$]   &  148                    &   fixed   \\
$D\,{\rm [pc]}$                &  161.8                  &   fixed   \\
\hline
\end{tabular}
\linespread{1.0}\selectfont
\tablefoot{(a) The effective temperature is derived using the spectral type-temperature conversion of \citet{kenyon1995}; see also similar 
measurements in \citet{mora2001} and \citet{alecian2013}. (b) The stellar luminosity and extinction are derived from fitting the optical 
SED with a photosphere model from \citet{Kurucz1994}.}
\label{tab:parameter}
\end{table}

%\subsection{Stellar heating}
%The disk is assumed to be passively heated by stellar irradiation. We take the Kurucz atmosphere model with ${\rm log}\,g=3.5$ as the incident 
%stellar spectrum \citep{Kurucz1994}. The effective temperature ($T_{\rm eff}$) and luminosity of the star ($L_{\star}$) were fixed to $8460\,\rm{K}$ 
%and $17.3\,L_{\odot}$ \footnote{The stellar luminosity $17.3\,L_{\odot}$ is higher than the value of ${\sim}13\,L_{\odot}$ derived before.
%This is because we use the GAIA-DR2 distance of $162\,\rm{pc}$ compared to $140\,\rm{pc}$ commonly adopted in previous work.}, respectively. 
%The radiative transfer problem is solved self-consistently considering 160 wavelengths, which are logarithmically distributed in the range of [$0.1\,\mu{\rm{m}}$, $1\,\rm{cm}$]. 

\subsection{The distribution of large grains}
\label{sec:surdens}

The features of the high-resolution ALMA observations are difficult to parameterize with a simple analytic expression, e.g., a power-law profile with a 
Gaussian perturbation at the location of the gap \citep[e.g.,][]{liu2017}. The surface density $\Sigma_{\rm LGP}$ is instead built by reproducing the 
brightness distribution of 1.3 mm continuum emission along the disk major axis, following an iterative procedure introduced by \citet{pinte2016}.

For a given surface density distribution, the radiative transfer code \texttt{RADMC-3D} is used to produce a synthetic image of 1.3\,mm continuum 
emission, which is then convolved with the beam of our ALMA observation.  The initial surface density profile is obtained from a functional form 
analogous to Eq.~\ref{eqn:sigma}, though the starting profile does not affect the final result. The distribution of the large grain population is 
truncated at $R_{\rm out,LGP}\,{=}\,200\,\rm{au}$, which is close to the boundary of emission from the disk shown in previous interferometric 
measurement at millimeter wavelengths \citep{pietu2006,hamidouche2006,guilloteau2011,huang2017}, as well as in our ALMA observation. The vertical 
structure of large dust grains is described with a functional form analogous to Eq.~\ref{eqn:dens}. To account for the effect of dust settling, 
large grains are less vertically distributed, with a scale height of $h_{\rm LGH}\,{=}\,{\Lambda}h_{\rm SGP}$ and a settling 
parameter $\Lambda\,{\leq}\,1$ constant over radius.

In order to keep the procedure as simple as possible, when constructing the surface density distribution we initially fix $H_{100}\,{=}\,10\,\rm{au}$ 
and $\beta\,{=}\,1.1$, as values that are typical of other disks \citep[e.g.,][]{andrews2009, madlener2012, lium2012, Kirchschlager2016}. The total 
dust mass in the disk $M_{\rm dust,tot}$ is initially fixed to $2\times10^{-3}\,M_{\odot}$, consistent with the results given by \citet{pietu2007} 
and \citet{guilloteau2011} assuming a gas-to-dust mass ratio of 100. The settling parameter $\Lambda=0.2$ and $f_{\rm LGP}=0.85$ are also fixed 
from previous models of protoplaneraty disks \citep[e.g.,][]{andrews2011,fang2017,fedele2018}. The initial inclination of $i=36^{\circ}$ is consistent 
with previous measurements \citep{simon2000,pietu2006,guilloteau2011}. 

The brightness distribution of the model is then compared to the radial brightness distribution extracted from the ALMA image along the disk major 
axis (see Figure~\ref{fig:obsflux}). In subsequent iterations, the large grain population surface density at each radius is corrected by the ratio 
of the observed and the modeled brightness profile. The iteration procedure converged after 30 iterations, when the change in model brightness profile 
was less than 5\% at all radii. The profile of the constructed $\Sigma_{\rm LGP}$, shown in the bottom panel of Figure~\ref{fig:surdens}, traces the 
gap and ring seen in the 1.3 mm continuum emission.

\begin{figure}[!t]
\includegraphics[width=0.48\textwidth]{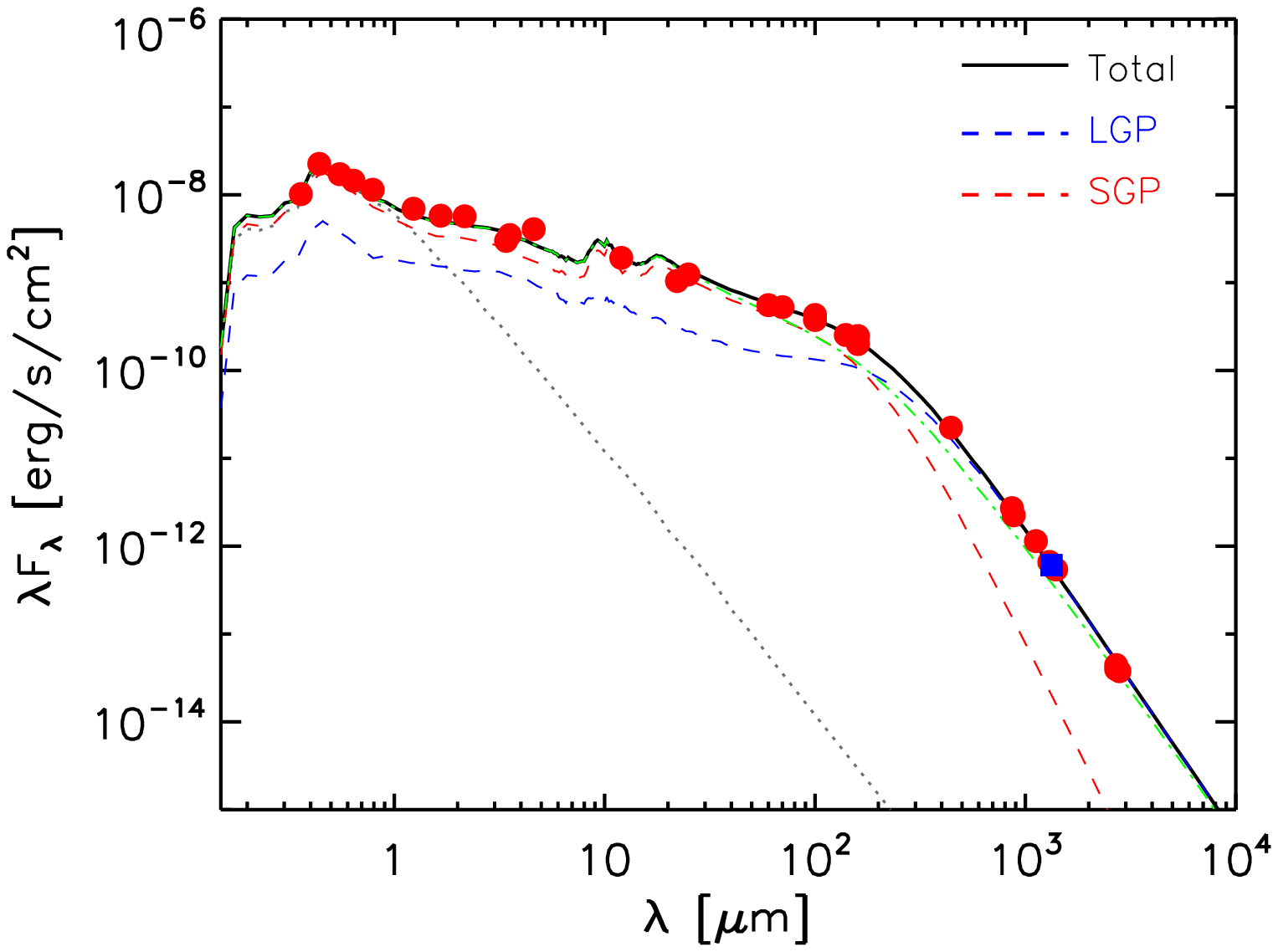}
\caption{The SED of the MWC 480 disk. The black line represents the best-fit model, while red dots are photometry collected from the 2MASS \citep{cutri2003}, 
WISE \citep{cutri2013}, IRAS \citep{beichman1988}, and AKARI surveys \citep{ishihara2010}, Herschel/PACS \citep{pascual2016}, {\it UBVRI} \citep{mendigutia2012}, 
and sub-mm bands \citep{mannings1997,hamidouche2006,pietu2006,hughes2008,oberg2010,guilloteau2011,huang2017}.  The flux density at ALMA Band 6 in this 
study is highlighted with a blue square. The contributions from the small grain population and large grain population to the total flux are indicated 
with the red and blue dashed lines, respectively. The green dash-dotted line gives the resulting SED when the raytracing only takes into account the 
central emission blob ($R_{\rm in} \le R \le 63\,\rm{au}$) as seen in our ALMA image, see Sect. \ref{sec:seddiscuss}. The gray dashed line is the 
input photosphere model \citep{Kurucz1994}.}
\label{fig:sedfit}
\end{figure}

The surface density profile obtained from this fit is then used as an input for a simultaneous fit to the SED and ALMA image by exploring the parameter 
space that was initially fixed in this subsection.

\subsection{Fit to the SED and brightness distribution}
\label{sect:fit}
After constructing the surface density profile for the large grain population, the next step of our modeling is to search for a coherent model that can 
explain the SED and the brightness distribution along the major and minor axis simultaneously. To achieve this goal, we explore the parameter 
space \{$M_{\rm dust,tot}$, $H_{100}$, $\beta$, $\Lambda$, $f_{\rm LGP}$, $i$\} using the simulated annealing algorithm \citep{kirkpatrick1983}.
Based on the Metropolis-Hastings algorithm, simulated annealing creates a Markov Chain Monte Carlo random walk through the parameter space, thereby 
gradually minimizing the discrepancy between observation and prediction by following the local topology of the merit function. 
This approach has specific advantages for high-dimensionality optimization because no gradients need to be calculated, and entrapment in local 
minima can be avoided regardless of the dimensionality. Details about the implementation of simulated annealing can be found in \citet{madlener2012} and \citet{liu2017}. 

In order to construct the observed SED with a complete wavelength coverage from optical to millimeter, we incorporate a variety of
archival data and multi-wavelength photometry. The IR excess from 1 to $10\,\mu{\rm m}$ of MWC 480 is known to vary by $\sim$30\% \citep{sitko2008, kusakabe2012, fernandes2018}. 
\citet{fernandes2018} explored two different possibilities: changing the height of the inner rim and utilizing a structure that simulates an inner disk wind. 
They found that both scenarios can reproduce the near-IR variability, but only the disk wind model can produce flux levels in the far-IR that are consistent 
with existing data. We do not attempt to re-investigate the same story given the fact that the ALMA data presented here are not sensitive enough to place 
additional constraints on the structure or geometry of the innermost region of the disk. The collected data points between 1 and $10\,\mu{\rm m}$ shown in 
Figure~\ref{fig:sedfit} are close to the average level of the observed fluxes. Our radiative transfer analysis is devised to derive disk characteristics in 
a more general way, with the focus of constraining the surface density and model parameters that describe the overall structure of the disk. These information 
are useful for the hydrodynamical simulation of planet-disk systems, see Sect.~\ref{sec:hydromodeling}.

\begin{figure}[!t]
\includegraphics[width=0.45\textwidth]{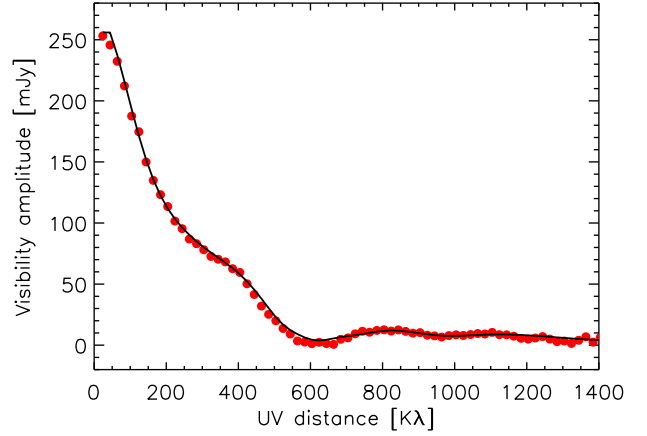}
\caption{Comparison between the observed (red dots) and model (black line) visibilities. The errors of the observation are smaller than the symbol size.}
\label{fig:visibility}
\end{figure}

The best-fit parameters are given in Table~\ref{tab:parameter}, with the results presented in Figures~\ref{fig:obsimg}, \ref{fig:obsflux}, \ref{fig:sedfit} 
and \ref{fig:visibility}. The best-fit model well reproduces the ALMA image and the SED. The scale height $H_{100}$ and total dust mass $M_{\rm dust,tot}$ 
are comparable to previous measurements for the MWC 480 disk \citep[e.g.,][]{pietu2006,pietu2007,guilloteau2011} and are typical for other protoplanetary 
disks \citep[e.g.,][]{madlener2012,grafe2013,Garufi2014,liu2017}. The relatively flat IR SED of MWC\,480 is consistent with being a member of the Meeus 
Group II \citep{meeus2001, sitko2008}. Therefore, a relatively small flaring index ($\beta\,{=}\,1.08$) appears preferable for the disk vertical geometry.

\begin{figure}[!t]
 \includegraphics[width=0.48\textwidth]{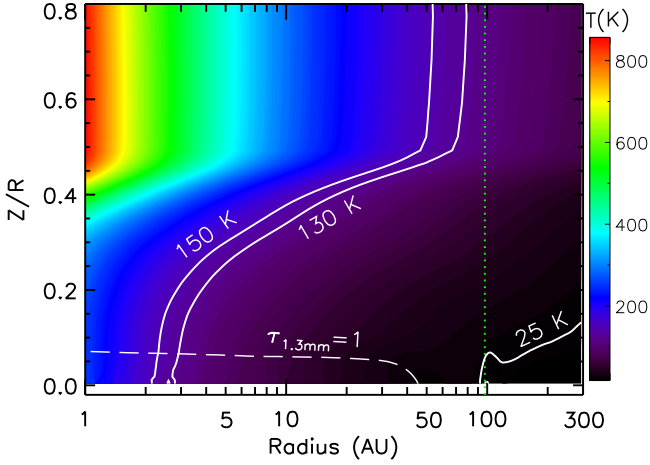}
 \caption{The 2D temperature structure of the best-fit radiative transfer model. Overplotted are contours at 25, 130 and $150\,\rm{K}$, which are 
relevant to the discussion on the origin of the ring structure (see Sect.~\ref{sec:gaporigin}). The vertical dotted line marks the center of the 
ring, i.e., $R\,{=}\,97.5\,\rm{au}$. The dashed curve indicates the surface of dust opacity being one ($\tau_{1.3\,{\rm mm}}\,{=}\,1$), where 
$\tau_{1.3\,{\rm mm}}$ is vertically integrated from the disk surface layer down to the midplane.}
\label{fig:temp2d}
\end{figure}

The ring has a dust mass of $2\times10^{-4}\,M_{\odot}$ (${\sim}67\,M_{\oplus}$), calculated by integrating the total surface density profile between 
81 and 117\,au, corresponding to the FWHM of the Gaussian fit to the brightness distribution in that ring. The ring mass of MWC\,480 is about half of 
the mass of the B7 ring (at a similar location of ${\sim}\,97\,\rm{au}$ from the star) in the HL Tau disk (see Fig.~3 in \citealt{alma2015}, \citealt{pinte2016,liu2017}). 
Figure~\ref{fig:temp2d} shows the 2D temperature structure of the best-fit model, in which contours at 25, 130 and $150\,\rm{K}$ are also marked. The 
temperature in the disk midplane generally follows a power law $T_{\rm midplane}\,{\propto}\,R^{-0.51}$ with $T{\sim}\,30\,\rm{K}$ at the gap location. 
The midplane temperature gradient is consistent with the results from previous parameter studies using a similar model setup \citep[e.g.,][]{pinte2014}.
 
\section{Modeling the dust gap with hydrodynamics gas/dust simulations}
\label{sec:hydromodeling}

Planet-disk interaction is an attractive mechanism to create gap-like structures in protoplanetary disks. To explore this hypothesis, we model the 
features of the dust continuum emission of the MWC\,480 disk by assuming that the observed gap (ring) corresponds to a dip (bump) in the dust density 
distribution of large grains (as shown in the bottom panel of Figure~\ref{fig:surdens}) induced by the presence of an embedded protoplanet.  Any structure 
detected in optically-thin disk emission may instead be caused by temperature effects or by changes in the optical properties of dust grains. In this 
work, we assume that these two contributions have a  negligible effect on the disk emission with respect to the effects of dust density.
For this reason, we assume a locally isothermal equation of state and an uniform choice for the shape, porosity and chemical composition of 
the dust across the disk.

Based on the result from the radiative transfer analysis described in Sect.~\ref{sec:rtmodeling}, we perform 3D simulations of a variety of dust/gas 
disk models hosting an embedded protoplanet. The dynamics of large dust grains yields an estimate of the main physical properties of a planet that 
could carve such a gap in the dust density distribution of millimeter grains. Given the large number of parameters involved in our analysis and the 
computationally expensive approach, the fitting procedure is done by eye without performing any statistical test to quantify the goodness of the fit. 

We perform a set of 3D Smoothed Particle Hydrodynamics (SPH) simulations of gas and dust disk with an embedded protoplanet using the \textsc{phantom} 
code \citep{price17a}. The dust disk is approximated as a two component system in terms of dust species: a population of micron-sized grains perfectly 
mixed with the gas and a population of millimetre-sized dust grains. The dynamics of dust grains are simulated using the one-fluid algorithm \citep{price15a} 
based on the terminal velocity approximation \citep[e.g.][]{youdin05a} and best suited to simulate particles tightly coupled to the gas \citep{ballabio18a}. 
We represent the protoplanet and the central star using sink particles \citep{bate1995}. The sinks are free to migrate and are able to accrete gas 
and dust particles due to their interaction with the disk. Particles are accreted onto the sinks when two conditions are fulfilled: i) the 
SPH particle is found to be gravitationally bound the sink, and ii) the divergence of the velocity field at the location of the particle 
is negative \citep{bate1995}.

Each simulation is evolved for a maximum time of 100 orbits at the initial planet distance from the central star. The gas and dust density distributions 
at the end of our simulations are not in a steady state. Since in our simulations the planet is allowed to migrate and accrete due to the tidal interaction 
with the surrounding disk, the planet properties and the tidal effect on the surrounding disks evolves with time without reaching a steady state. It is 
therefore reasonable to expect a degree of degeneracy with the simulation parameters (e.g. initial planet mass, location, gas disk mass). However, given 
the large range of parameters involved in the fitting procedure, we cannot reasonably quantify this degeneracy, limiting our analysis to a set of initial 
disk and planet properties that are able to reproduce the dust disk model of MWC 480, as obtained from the radiative transfer modeling.

\subsection{Initial conditions}
Our reference model for the total dust mass, grain size distributions, and the disk structural properties are adapted from the results of the 
radiative transfer modeling described in Sect.~\ref{sec:rtmodeling}. The system consists of a central star of mass $M_{\star}$ (see Table~\ref{tab:parameter}) 
surrounded by a gas and dust disk extending from $R_{\mathrm{in}}\,{=}\,1\,\rm{au}$ to $R_{\mathrm{out}}\,{=}\,150\,\rm{au}$ and modeled as a set of $5\,{\times}\,10^{5}$ 
SPH gas/dust particles. To simplify our analysis, gas and dust are assumed to be initially well mixed across the disk and the dust and gas surface density 
profile are initially assumed to be power-laws with the same exponential index ($\gamma=1$), without the tapering of Eq.~\ref{eqn:sigma}. However, our 
simulations show that the dynamics of large grains naturally produces a tapering of the dust surface density in the outer disk region. Therefore, the 
choice for the shape of the initial dust surface density in the outer disk does not have a significant impact on the results at the end of the simulation.

The disk is vertically extended by assuming Gaussian profiles for the volume density in the vertical direction with a thickness expressed by 
Eq.~\ref{eq:heightgas}. We adopt a vertically isothermal equation of state $P=c_{\rm s}^{2} \rho_{\mathrm{g}}$ with $c_{\rm s}\left(R\right) = c_{{\rm s,in}} (R/R_{\rm in})^{-q}$ 
with a value $c_{{\rm s,in}}$ and $q$ computed by assuming vertically hydrostatic equilibrium across the disk, i.e. $H_{\mathrm{g}}=c_{\rm s}/\Omega_{\mathrm{k}}$.
Using the model parameters shown in Table~\ref{tab:parameter}, we obtain $H_{\mathrm{g}}/R_{\mathrm{in}} =0.083$ and $q=3/2-\beta=0.42$. We set the 
SPH $\alpha_{\rm AV}$ viscosity parameter to $\sim$0.15, resulting in an effective \citet{shakura73a} viscosity $\alpha_{\mathrm{SS}} \approx 5 \times 10^{-3}$.

In order to explore different dust density structure resulting from different gas and planet properties, we perform a suite of simulation by varying the 
total dust-to-gas ratio and the planet properties (mass and initial separation).

\begin{figure*}
\begin{center}
\includegraphics[width=0.35\textwidth]{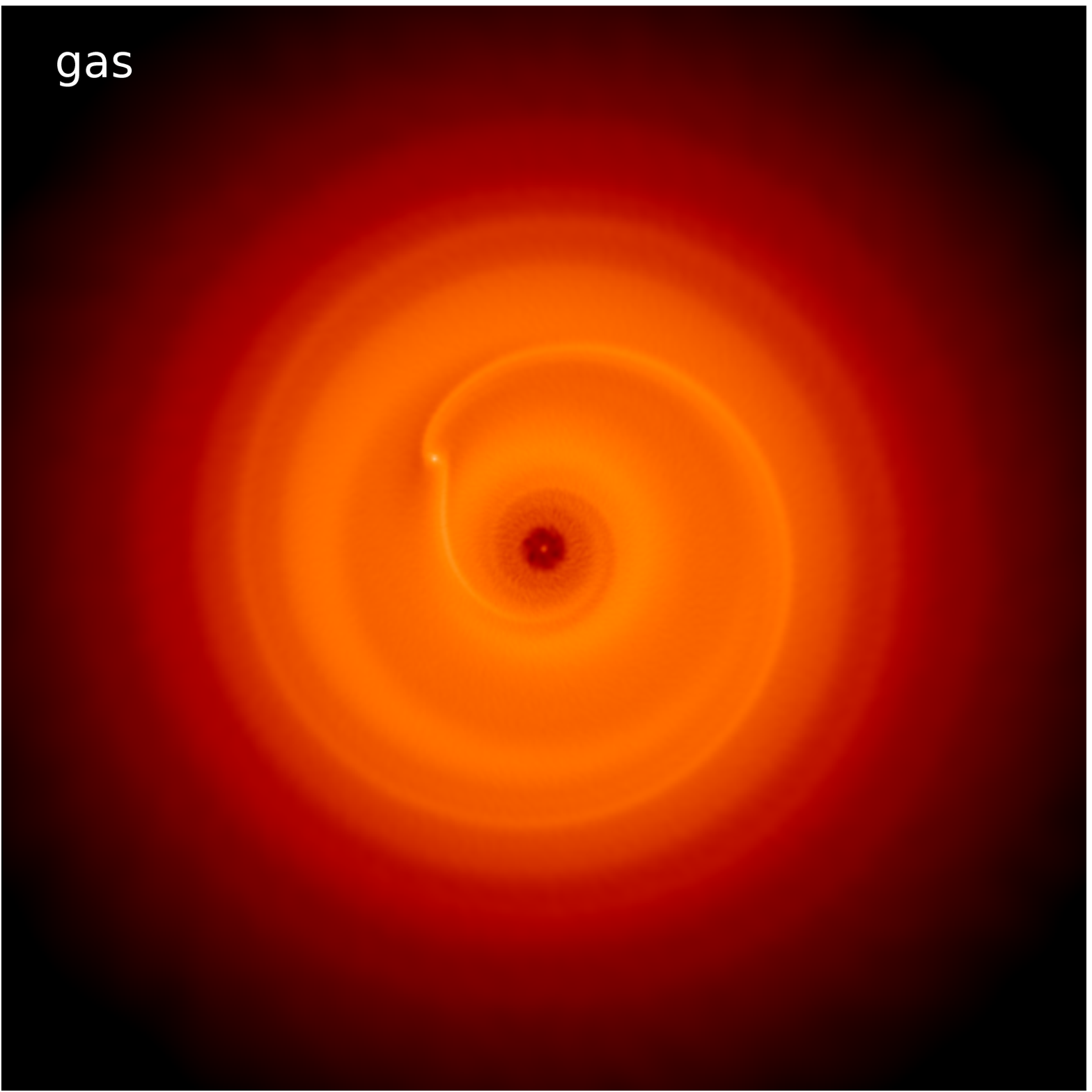}
\includegraphics[width=0.35\textwidth]{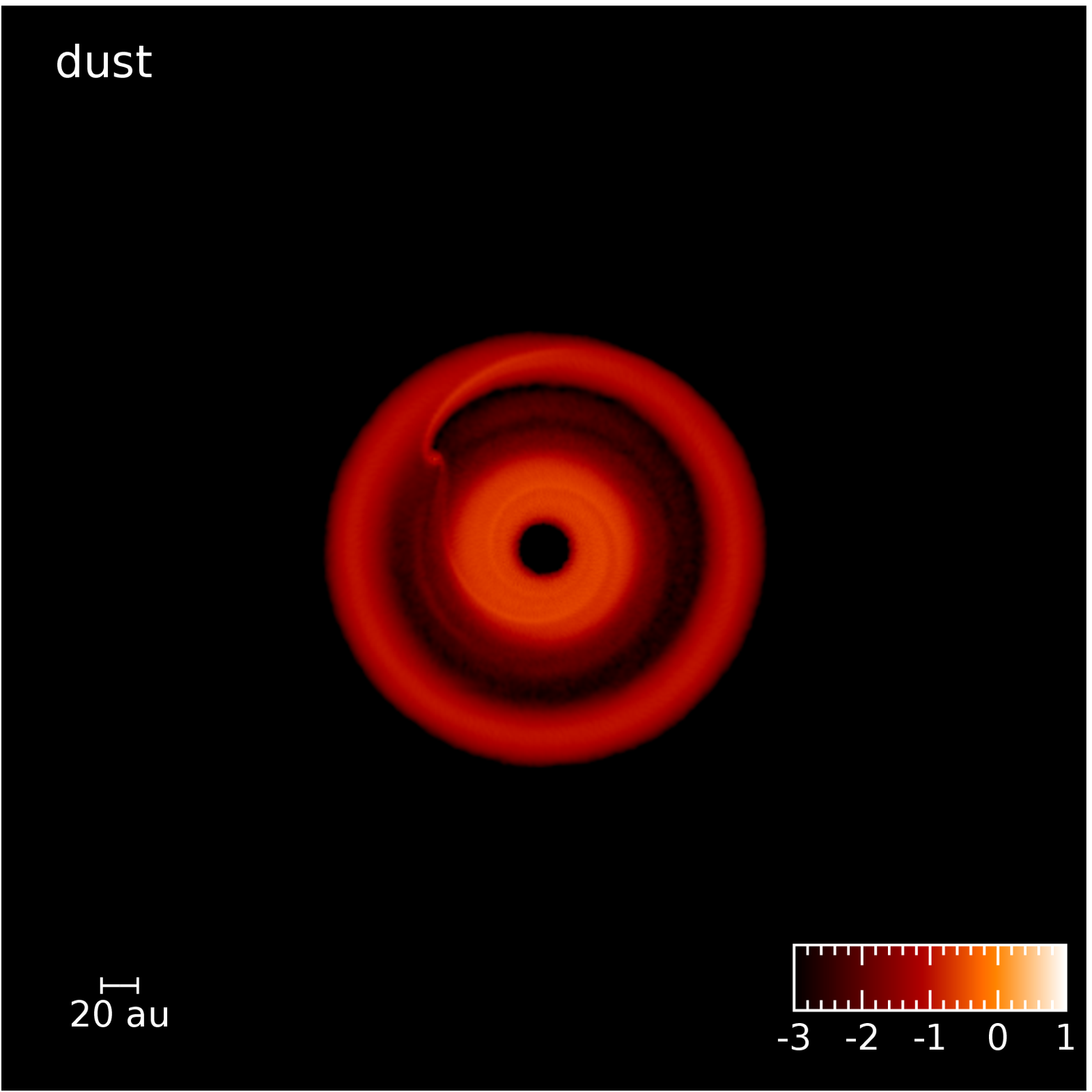}
\caption{Rendered images of gas (left) and millimetre dust grain (right) surface density (in units of $\mathrm{g\,cm^{-2}}$ on a logarithmic scale) of 
our best-fit disk model hosting a planet with an initial mass of $1.25 \, \mathrm{M_{\mathrm{J}}}$ and initially located at a distance of 85 au from the central star.}
\label{fig:renderingsigmagd}
\end{center}
\end{figure*}

\subsubsection{Dust-to-gas mass ratio}
One of the key parameters in disk modeling is the gas disk mass co-located with the dusty disk emission shown in the ALMA image. The local gas surface 
density regulates the aerodynamical coupling between the gas and the dust \citep[e.g.][]{weidenschilling77b} and the accretion and migration efficiency 
of the planet \citep{dangelo08a,dangelo10a}. Aerodynamical coupling is usually described in terms of the Stokes number St:
\begin{equation}
\rm{St}= t_{\rm stop} \Omega_{\mathrm{k}}\approx \frac{\rho_{\rm d}{\rm a}}{\Sigma_{\rm gas}},
\end{equation}
where $t_{\rm stop}$ is the aerodynamical stopping time, $\Omega_{\mathrm{k}}$ is the Keplerian angular velocity, $\rho_{\rm d}$ is the internal 
dust density and $a$ is the grain size. 

The typical approach is to assume a fixed value throughout the entire disk given by the typical ISM dust-to-gas mass ratio $M_{\rm dust}/M_{\rm gas}=0.01$. 
However, the small grains (and thus the gas) have been observed to extend up to $R_{\rm out,SGP}\,{\sim}\,750\,\rm{au}$ while the millimeter-sized grains 
extend up to $R_{\rm out,LGP}\,{\sim}\,200\,\rm{au}$ (Figure~\ref{fig:surdens}). Assuming that originally the two dust populations had the same radial 
distribution, and that the discrepancy between them is only due to the radial drift of large grains \citep[e.g.][]{birnstiel14a}, it is reasonable to 
believe that the dust-to-gas mass ratio has increased by a factor ${\sim}\,\left(R_{\rm out,LGP}/R_{\rm out,SGP}\right )^2\,{\sim}\,14$ with respect 
to the value at birth (i.e. $M_{\rm dust}/M_{\rm gas}\,{\approx}\,0.01$), in the regions of the MWC\,480 disk where millimeter-sized grains reside. 
Therefore, our simulations use dust-to-gas mass ratios that span the range $M_{\rm dust}/M_{\rm gas}\in [0.01,0.2]$. Given the best-fit total dust 
mass $M_{\rm dust,tot}$ from the radiative transfer modeling, this choice of $M_{\rm dust}/M_{\rm gas}$ corresponds to a disk gas mass 
between $0.008\, {\rm M_\odot}\,{<}\,M_{\rm disk,g}\,{<}\,0.16\, {\rm M_\odot}$.

Adopting the ISM value $M_{\rm dust}/M_{\rm gas}\,{=}\,0.01$ would have yielded a gas disk that is close to be gravitationally unstable, as assessed 
by the stability criterion of \citet{toomre64a} of $Q\equiv c_{\mathrm{s}} \Omega_{\mathrm{k}}/\pi \mathcal{G} \Sigma_{\mathrm{g}}$, where $\mathcal{G}$ 
is the gravitational constant. In the outer disk, the value of the $Q$-parameter would reach values close to 3, which implies that the disk would be 
marginally prone to be gravitational unstable \citep{kratter16a}. In this case, the disk would show the development of large-scale density fluctuations 
in the form of a spiral pattern \citep[e.g.][]{cossins09a} that would have been observed by our observations \citep[e.g.][]{dipierro15a}. Recent ALMA 
surveys also found a dust-to-gas mass ratio higher than the ISM value in many protoplanetary disks \citep[e.g.,][]{ansdell16a,ansdell2017,long2017}. 
Therefore, it is reasonable to assume $M_{\rm dust}/M_{\rm gas}\,{>}\,0.01$. 

In our simulations, millimeter-sized dust particles are characterized by a Stokes number in the range ${\sim}\,[0.001,0.1]$, with a value 
of $\mathrm{St}\,{\sim}\,0.05$ at the planet location for $M_{\rm dust}/M_{\rm gas}\,{=}\,0.1$. The value of the Stokes number in the outer disk 
affects the shape of the dust surface density of millimeter grains. As shown in the bottom panel of Figure~\ref{fig:surdens}, the ring structure is 
just outside the gap, which suggests that the ring might be a trap of millimeter-sized grains due to the presence of a pressure maximum at the outer 
edge of the gap carved by the planet. The efficiency of dust trapping and, thus, the shape of the dust surface density in the outer disk region 
depends on the local gas properties. Moreover, that shape can be also explained by the drift-dominated dynamics of non-growing grains 
with $\mathrm{St}\,{<}\,1$ \citep{youdin02a,birnstiel14a} migrating from the outer disk toward the planet location (see Sect.~3.3.1 in \citealt{dipierro2018}). 
Analyzing dust settling of large dust grains constrained via radiative transfer modeling (see Sect.~\ref{sec:rtmodeling}) can yield a rough estimate 
of the gas disk mass. The uncertainty of the result is connected with how well $\alpha_{\mathrm{SS}}$ is constrained.
In detail, the vertical balance between the gravitational settling and turbulent diffusion determines the thickness of the dust 
disk, given by \citep{dubrulle95a,fromang09a}

\begin{equation}
\frac{H_{\mathrm{d}}}{H_{\mathrm{g}}} = \sqrt{\frac{\alpha_{\mathrm{SS}}}{\alpha_{\mathrm{SS}}+\mathrm{St}}}.
\end{equation}
Assuming $\alpha_{\mathrm{SS}}\,{=}\,5\,{\times}\,10^{-3}$ and $\mathrm{St}$ computed by assuming $M_{\rm dust}/M_{\rm gas}\,{=}\,0.1$, we 
obtain a dust-to-gas height ratio ${\sim}\,0.28$, averaged over a distance from the central star $R\,{\in}\,[30\, \mathrm{au} ,R_{\mathrm{out}}]$. 
This value is quite close to the value of $\Lambda\,{=}\,0.25$ (see Table~\ref{tab:parameter}) inferred from the radiative transfer modeling, implying 
that the dust vertical structure generated by a dust-to-gas mass ratio of $0.1$ would reproduce the expected level of dust settling.

\subsubsection{Analytical estimate of the planet properties}
\label{sect:analplanet}

We run simulations adopting different planet mass and radial distance from the central star. Due to the uncertainties in the planet properties, we 
consider a range of planet masses and distances based on analytical estimates of their fiducial values.

Numerical simulations have shown that the width $\Delta$ of dust gaps opened by planets is a few times the Hill radius $\Delta \approx x R_{\rm Hill}$ 
where $x=4\textrm{--}8$ \citep{rosotti2016}, and 

\begin{equation}
R_{\rm Hill}=\left(\frac{M_{\rm p}}{3M_\star}\right)^{1/3}R_{\rm p},
\end{equation}
where $R_{\rm p}$ is the radial distance of the planet. Therefore, we can qualitatively estimate the planet mass producing the dust density 
depletion seen in our ALMA continuum data as
\begin{equation}
M_{\rm p}\approx 3\left( \frac{\Delta}{xR_{\rm p}} \right)^3 M_\star\sim 0.4 \textrm{--}3\, {\rm M_J},
\end{equation} 
assuming $R_{\rm p}=73.4 \,{\rm au}$ and $\Delta=23 \,{\rm au}$.

This range of planet masses is consistent with the minimum planet mass required in order to carve a gap in the dust density profile \citep{rosotti2016,dipierro2017,facchini2018}
\begin{equation}
M_{\rm p}\approx \min\left[0.3\left(\frac{H}{R}\right)^3_{R_{\rm p}}, 1.38\left(\frac{\zeta}{\rm St}\right)^{3/2}\left(\frac{H}{R}\right)_{R_{\rm p}}^3 \right]M_\star\sim 1 {\rm M_J},
\end{equation}
where
\begin{equation}
\zeta=-\left.\frac{\partial  \ln P}{\partial \ln R}\right|_{R_{\rm p}}=\left(\gamma+1.5-\beta+\frac{3}{2}\right)=2.92,
\end{equation}
using the parameters shown in Table~\ref{tab:parameter}. Recently, \citet{ataiee2018} found the following criterion to produce a pressure 
bump able to trap dust grains as 
\begin{equation}
M_{\rm p}\gtrsim \left(\frac{H}{R}\right)_{R_{\rm p}}^{3}M_\star \sqrt{82.33\alpha_{\rm SS}+0.03}\sim 2 \,{\rm M_J}.\label{ataieecrit}
\end{equation}
Based on this analysis, our numerical simulations adopt planet masses spanning $M_{\rm p}=[0.4,2] \, \rm{M_{J}}$.

%The case ``inertial'' criterion, for fast migrating satellites which reads \citep{ward1989}
%\begin{equation}
%M_{\rm p}\gtrsim \frac{\Sigma R_{\rm p}^2}{M_\star} \left(\frac{H}{R}\right)_{R_{\rm p}}^3M_\star.
%\end{equation}

We now check whether the masses inferred from the previous analysis are expected to carve a deep gap also in the gas. In order to open 
a deep gap (corresponding to a depletion of $\sim 90\%$ of the unperturbed value), a planet needs to satisfy the following criterion \citep{crida2006}
\begin{equation}
\frac{3}{4}\left(\frac{H}{R}\right)_{R_{\rm P}}\left(\frac{M_{\rm p}}{3M_\star}\right)^{-1/3}+\frac{50 \nu}{\Omega_{\rm P}R_{\rm P}^2}\left(\frac{M_{\rm p}}{M_\star}\right)^{-1}\lesssim 1,
\label{cridacrit}
\end{equation}
where $\Omega_{\rm p}=2{\rm \pi}/t_{\rm p}$ is the Keplerian orbital frequency of the planet, and $t_{\rm p}$ its orbital period. 
The criterion in Eq.~\ref{cridacrit} requires a satellite with $M_{\rm p}\gtrsim 16 \,{\rm M_{\rm J}}$ in order to open a deep gap in the gas. 
This result suggests that the planet masses simulated here will carve a gap in the dust, and will produce only a slight 
depletion ($10\,\textrm{--}\,20 \%$) in the gas density at the planet location \citep{ataiee2018}.

As soon as the simulation begins, the planet starts migrating inward as a consequence of the interaction with the disk. Since the planet 
is initially completely embedded in the disk, it migrates according to type I migration on a timescale from \citet{paardekooper2010} of
\begin{equation}
t_{\rm typeI}=\frac{R_{\rm p}}{\dot R_{\rm p}} \approx \left(\frac{M_{\rm p}}{M_\star}\right)^{-1}\frac{M_\star}{\Sigma_{R_{\rm p}}R_{\rm p}^2}\left(\frac{H}{R}\right)_{R_{\rm P}}^{2}\Omega_{\rm p}^{-1}\sim 30\,\textrm{--}\,600 \,t_{\rm p},
\label{eq:type1migr}
\end{equation}
where $\Sigma_{R_{\rm p}}$ is the gas surface density at the planet radial distance from the star. On the right hand side of 
Eq.~\ref{eq:type1migr}, the shortest timescale is the one for $M_{\rm dust}/M_{\rm g}=0.01$ and the longest is the one for $M_{\rm dust}/M_{\rm g}=0.2$.

Since the planet should migrate inward due to the planet-disk interaction in our disk model, we initially locate the planet on an outer 
orbit ($R_{\rm p}=85\,\textrm{--}\,120\,\rm{au}$, depending on the migration timescale) with respect to the position of the minimum 
of the intensity profile in the gap region (i.e. $R_{\mathrm{p}}\,{\gtrsim}\,74.3$\,au). 

\begin{figure}
\includegraphics[width=0.48\textwidth]{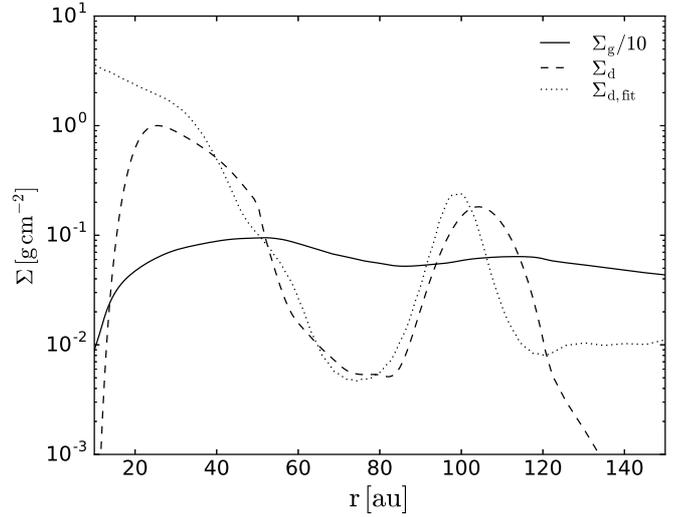}
\caption{Azimuthally averaged dust (dashed) and gas (solid) surface density for our best disk hosting a planet with an initial mass 
of $1.25 \,\mathrm{M_{\mathrm{J}}}$ after 50 orbits, initially located at 85\,au from the central star. The gas surface density is scaled 
down by a factor of 10 at the end of the simulation, for a direct comparison with the dust phase. The dotted line indicates the surface 
density profile of large dust grains inferred from the radiative transfer analysis (bottom panel of Figure~\ref{fig:surdens}). After 50 
orbits (measured at 85\,au), the planet is at 78.4\,au from the central star and its mass is $2.3\, \mathrm{M_{\mathrm{J}}}$.}
\label{fig:besthydro}
\end{figure}

\subsection{Results}

Our best model consists of an initial well-mixed gas and dust disk with $M_{\rm dust,tot}/M_{\rm gas}\,{=}\,0.1$, hosting a planet with 
an initial mass of $1.25\, \mathrm{M_{\mathrm{J}}}$ located at a distance of 85\,au from the central star. We stop the simulations 
after ${\sim}\,50$ orbits at the initial location of the planet. Figure~\ref{fig:renderingsigmagd} shows the rendered images of gas and 
millimeter dust grain surface density of our best-fit model. The surface density structure of the dust phase shows an annular gap around 
the planet location. In contrast, the gas phase shows a weak depletion asymmetric with respect to the planet location and a spiral 
structure across the disk. The fast radial drift of dust grains in the outer disk toward the inner disk regions produces the accumulation 
of dust grains at the gap outer edge. After 50 orbits (measured at 85\,au), the planet reaches the distance of 78.4\,au from the central 
star and its mass is ${\sim}\,2.3\,\mathrm{M_{\mathrm{J}}}$ with an average mass accretion rate 
of ${\sim}\,3 \times 10^{-5} \,\mathrm{\mathrm{M_{\mathrm{J}}}\,yr^{-1}}$. Moreover, the star accretes mass from the surrounding gas and 
dust disk at a rate given by ${\sim}\,2.5 \times 10^{-7} \,\mathrm{\mathrm{M_{\odot}}\, yr^{-1}}$. This value is consistent with the average 
stellar accretion rate of  $1.2\times10^{-7}$\,M$_\odot\,\rm{yr}^{-1}$ measured by \citet{mendigutia2013} \footnote{The original value reported 
in the paper has been corrected to nearly the same value by accounting for the updated distance, stellar mass, and stellar radius adopted in our analysis.}.
It is worth remarking that the fast planet migration and accretion can be slowed down in our model by running the simulation for 
longer, letting the planet carve a deep gas gap.

Figure~\ref{fig:besthydro} shows the azimuthally averaged surface density of the gas and dust at the end of the simulation, compared to the 
fit for the dust surface density of large grains described in Sect.~\ref{sec:rtmodeling}. As expected, the gap in the dust component is 
deeper than the (barely visible) gap in the gas due to both the higher efficiency of the tidal torque and the dust radial motion induced 
by drag \citep{paardekooper06a,fouchet07a,pinilla15c,dipierro2017}. Our simulation shows that the planet perturbs the local pressure profile 
and create a pressure maximum at the gap outer edge. As a result, millimeter grains accumulate at the location of the pressure maximum, forming 
a deeper dust gap than in the gas. At the inner edge, the perturbation of the pressure profile induced by the tidal torque does not exceed 
the background pressure gradient, leading to a weak pile up at the inner gap edge and an accelerated radial inflow toward the central 
star \citep{fouchet10a}. The discrepancy in the density profiles with respect to the radiative transfer model inside ${\sim}\,20$\,au is 
most likely a numerical effect, due to the resolution-dependent SPH numerical viscosity, which causes an artificially fast depletion of 
the innermost disk regions \citep{lodato10a}.

\section{Discussion}
\label{sec:discussion}

\subsection{Why the SED does not show evidence of the gap}
\label{sec:seddiscuss}
The SED serves as an important diagnostic for substructures, including large inner holes (or cavities) and opacity gaps in protoplanetary disks. 
The significant depletion of small dust grains inside the hole and gap leads to a clear flux deficit in the near- and/or mid-IR wavelength domain
\citep[e.g.,][]{espaillat2007,espaillat2010,williams2011,espaillat2014}. High-resolution (sub-)millimeter observations confirm the presence of such holes and gaps 
that are indirectly inferred from the SEDs \citep[e.g.,][]{andrews2011}, and also yield hole sizes that are generally match well with estimates 
from SED modeling \citep{vandermarel2016}.

However, the SED of MWC\,480 does not show any significant decrease in the near- or mid-IR wavelength regime (see Figure~\ref{fig:sedfit}), even 
though our ALMA observation clearly reveals a large and deep gap in the disk. This apparent discrepancy can be explained by two reasons. First, 
the 1.3\,mm continuum observation is sensitive only to millimeter dust grains and does not probe the distribution of small grains in the disk. 
The polarimetric differential imaging in the H band obtained with Subaru/HiCIAO indicates that small dust grains are smoothly distributed up 
to ${\sim}160\,\rm{au}$ \citep{kusakabe2012}. These small grains are efficiently heated and their thermal re-emission contribute a vast majority 
of the observed flux at IR wavelengths (see the red dashed line in Figure~\ref{fig:sedfit}). Second, the gap is too far from the central star 
to have a significant impact on the IR SED. Based on our ALMA observations, the disk can be generally divided into three parts, i.e., a central 
bright blob, a gap and a bright ring. The green dash-dotted line in Figure~\ref{fig:sedfit} shows the SED of the central bright blob defined with 
a radius range of $R_{\rm in}\,{\le}\,R\,{\le}\,63\,\rm{au}$. Subtracting the gap location with half of the gap width gives $63\,\rm{au}$ as the 
outer radius. The central blob is already sufficient to shape the SED up to a wavelength of ${\sim}50\,\mu{\rm m}$. However, most 
criteria (e.g., using color-color diagram) for selecting disks with inner holes or gaps rely on data shorter than ${\sim}24\,\mu{\rm m}$ \citep[e.g.,][]{fang2009,cieza2010,merin2010}. 
Quantifying the impact of gaps located at large radii on the appearance of the SED would require a detailed parameter study that covers a 
broad range of gap location, width and depth, and also considers multiple gaps in a disk. 

A few disks have also been found to have inner holes in millimeter imaging without evidence for IR dips in their SEDs, including 
MWC 758 \citep{isella2010b,boehler2018,dong2018}, RY Tau \citep{isella2010a,pinilla2018,long2018}, and 
several disks in the Ophiuchus star-forming region \citep{andrews2011}.

\subsection{On the origin of the gap/ring structure}
\label{sec:gaporigin}
In Sect.~\ref{sec:hydromodeling}, we explored planet-disk interaction as the origin of the gap/ring structure in the MWC\,480 disk. The 
hydrodynamics simulations suggest that the observed gap is caused by gravitational perturbations from a planet with a mass of $2.3\,\rm{M_{J}}$ 
embedded in the gap. It is interesting to check whether the disk, particularly in the vicinity of the gap region, had sufficient material to 
form such a planet. For this purpose, we first performed a power-law fit to the best-fit surface density profile (the bottom panel 
of Figure~\ref{fig:surdens}) outside the gap. The result is assumed to be the initially unperturbed surface density profile of large 
dust grains. We integrated this power-law fitted profile within the gap region and derived an initial mass of the material. By integrating 
the best-fit surface density profile in the same radius range, we obtained an after-perturbed mass. The difference between these two 
integrated masses is ${\sim}5\,\rm{M_{J}}$, which can be considered as the ``missing material'' in the gap during the process of planet 
formation. This calculation uses a dust-to-gas mass ratio of 0.1 from the best hydrodynamical model. An assumption of a standard 
dust-to-gas mass ratio of 0.01 (i.e., for the interstellar medium), which is believed valid at the early stage of disk evolution, would 
increase the missing mass. This simple analysis indicates that there is a possibility for the formation of the planet.

Using polarimetric differential imaging in the H band obtained with Subaru/HiCIAO, \citet{kusakabe2012} found that the radial polarized 
intensity profile drops as $R^{-2}$ from $0.2^{\prime\prime}$ to $0.6^{\prime\prime}$ (32$-97$\,au), while it is steeper, $R^{-3}$, in the 
outer disk from $0.6^{\prime\prime}$ to $1.0^{\prime\prime}$ (97$-162$\,au). The authors compared the outer disk properties of MWC 480 with 
the roughly coeval SAO\,206462 and the younger LkCa 15, both of which share similar characteristics in the radial polarized intensity profile. 
The evidence of spiral arms in SAO\,206462 \citep{muto2012,maire2017} and planet candidates in LkCa 15 \citep{kraus2012,sallum2015} drive the 
authors to suggest that the outer disks in these systems are dynamically excited by massive planets. Our hydrodynamics modeling coupled with 
detailed radiative transfer analysis for MWC\,480 supports this hypothesis. Nevertheless, deep high-contrast searches for planets in the 
gap are required to confirm the presence of a planet \citep[e.g.,][]{testi2015,ruane2017,keppler2018}.

Alternative mechanisms besides planets can also create gap/ring structures in protoplanetary disks. 
Dust coagulation/fragmentation \citep{zhang2015,banzatti2015,pinilla2017} and sintering-induced dust rings \citep{okuzumi2016} near condensation 
fronts are believed to be universal, though whether they can produce gaps is unclear. In cold disk regions, dust grains are expected to be covered 
by ice mantles of volatile molecules. The condensation fronts, or known as snow lines, are defined as the location where the ice mantles evaporate 
during the inward drift of dust grains. Phase transition occurs when the dust grains cross the condensation fronts relative to the molecules that 
form their ice mantles, which alters the opacity, coagulation and fragmentation properties of dust grains, leading to the accumulation of dust 
particles in ring structures. In Figure~\ref{fig:temp2d}, we draw temperature contours of 25, 130, 150\,K corresponding to the condensation 
temperatures\footnote{The condensation temperature of volatiles can vary under different assumptions for the composition and surface area of 
the dust grains and the gas number density in the disk midplane. For instance, \citet{oberg2011} suggested an average condensation temperature 
of 20\,K for CO ice, while \citet{zhang2015} adopted a higher value of 25\,K.} of CO and $\rm{H_2O}$ that are abundant and commonly observed 
in protoplanetary disks. The center of the ring marked with the vertical dotted line appears associated with the CO snow line. In the disk 
midplane, the $\rm{H_2O}$ snow line is quite close to the central star and beyond the spatial resolution of our ALMA observation. One of the 
key feature of the condensation front as the origin of ring structure is that the total gas surface density is not expected to have strong 
variations \citep{pinilla2017,stammler2017}. However, gas observations towards the MWC\,480 disk so far lack sufficient spatial resolution 
to confirm a small variation in the gas surface density near the CO snow line. Future high-resolution observations of the gas 
components are highly desired to constrain the gas surface density profile of the MWC\,480 disk \citep[e.g.,][]{isella2016}.

The dead zone is another potential cause for the ring structures \citep{pinilla2016}. Protoplanetary disks embedded in a weak magnetic field 
are expected to be subject to the magnetorotational instability, which induces MHD turbulence to transfer angular momentum outward and therefore 
sustain accretion onto the  central star \citep{balbus1991,balbus1998}. The low degree of gas ionization in the dead zone suppresses the 
magnetorotational instability. Consequently, accretion slows down and a gas pressure bump is formed at the outer edge of the dead 
zone \citep{flock2015}. The pressure bump is capable of trapping particles and enhancing grain growth, leading to ring-like structures. 
In addition, the magnetized disk model has been supported further from comparisons between simulation and observation \citep{flock2017, bertrang2017}. 
Follow-up observations are required to investigate whether dead zones are responsible for the ring observed in the MWC\,480 disk. 
For example, spatially-resolved measurements of the turbulence through different layers in the disk  \citep[e.g.,][]{flaherty2015,teague2016} will help to test the size of dead zones.

\section{Summary}
\label{sec:summary}
In this work, a gap and bright dust ring are identified in the MWC\,480 disk for the first time using high-resolution ($0.17^{\prime\prime}\times0.11^{\prime\prime}$) 
ALMA band 6 continuum observations. The gap is located at $74.3\,\rm{au}$ from the central star and has a width of $23\,\rm{au}$. The ring is centered at $97.5\,\rm{au}$. 
In order to constrain the surface density and structural parameters of the disk, we performed detailed radiative transfer modeling, in which two grain 
populations (i.e., a small and large grain population) with different spatial distribution and dust properties are considered in the setup. Both the ALMA 
image at 1.3\,mm and the broadband SED are taken into account in the modeling procedure. 

Planet-disk interactions are one of the most attractive mechanisms for the formation of gap/ring structures. For MWC 480, we test this scenario by 
performing global three-dimensional SPH gas/dust simulations of disks hosting a migrating and accreting planet. The dust emission across the disk is 
consistent with the presence of an embedded planet with a mass of ${\sim}\,2.3\,\mathrm{M_{\mathrm{J}}}$ at an orbital radius of 
${\sim}\,78\,\rm{au}$. Given the surface density of the best-fit radiative transfer model, the amount of depleted mass in the gap is higher 
than the planet mass, which means that the disk had the potential to form such a planet at this location. Therefore, the MWC\,480 disk deserves 
follow-up deep and high-contrast search for planets. We also briefly discussed condensation fronts and dead zones as the origin of the gap/ring structure. 
Future high-resolution observations of the gas and multi-wavelength measurements of the depth and width of the gap are indispensable to identify 
the dominant mechanism that carve out gaps in disks.

\acknowledgements

YL acknowledges supports by the Natural Science Foundation of Jiangsu Province of China (Grant No. BK20181513) and by the Natural Science Foundation of China (Grant No. 11503087). 
GD acknowledges financial support from the European Research Council (ERC) under the European Union's Horizon 2020 research 
and innovation programme (grant agreement No 681601). GL and BN thank the support by the project PRIN-INAF 2016 The Cradle of Life - GENESIS-SKA (General 
Conditions in Early Planetary Systems for the rise of life with SKA). GJH is supported by general grants 11473005 and 11773002 awarded by the National 
Science Foundation of China. DH is supported by European Union A-ERC grant 291141 CHEMPLAN, NWO and by a KNAW professor prize awarded to E. van Dishoeck. 
YB, FM and GvdP acknowledge funding from ANR of France under contract number ANR-16-CE31-0013. DJ is supported by NRC Canada and by an NSERC Discovery Grant. 
CFM acknowledges an ESO Fellowship. This research used the ALICE High Performance Computing Facility and the DiRAC Data Intensive service operated 
by the University of Leicester IT Services. These resources form part of the STFC DiRAC HPC Facility (\url{www.dirac.ac.uk}), jointly funded by 
STFC and the Large Facilities Capital Fund of BIS via STFC capital grants ST/K000373/1 and ST/R002363/1 and STFC DiRAC Operations grant ST/R001014/1.

This paper makes use of the following ALMA data: 2016.1.01164.S. ALMA is a partnership of ESO (representing its member states), NSF (USA) and 
NINS (Japan), together with NRC (Canada), MOST and ASIAA (Taiwan), and KASI (Republic of Korea), in cooperation with the Republic of Chile. 
The Joint ALMA Observatory is operated by ESO, AUI/NRAO and NAOJ. This work has made use of data from the European Space Agency (ESA) mission 
Gaia (https://www.cosmos.esa.int/gaia), processed by the Gaia Data Processing and Analysis Con- sortium (DPAC, https://www.cosmos.esa.int/web/gaia/dpac/consortium). 
Funding for the DPAC has been provided by national institutions, in particular the institutions participating in the Gaia Multilateral Agreement.

\bibliographystyle{aa}
\bibliography{mwc480}

\begin{thebibliography}{144}
\expandafter\ifx\csname natexlab\endcsname\relax\def\natexlab#1{#1}\fi

\bibitem[{{Alecian} {et~al.}(2013){Alecian}, {Wade}, {Catala}, {Grunhut},
  {Landstreet}, {Bagnulo}, {B{\"o}hm}, {Folsom}, {Marsden}, \&
  {Waite}}]{alecian2013}
{Alecian}, E., {Wade}, G.~A., {Catala}, C., {et~al.} 2013, \mnras, 429, 1001

\bibitem[{{ALMA Partnership} {et~al.}(2015){ALMA Partnership}, {Brogan},
  {P{\'e}rez}, {Hunter}, {Dent}, {Hales}, {Hills}, {Corder}, {Fomalont},
  {Vlahakis}, {Asaki}, {Barkats}, {Hirota}, {Hodge}, {Impellizzeri}, {Kneissl},
  {Liuzzo}, {Lucas}, {Marcelino}, {Matsushita}, {Nakanishi}, {Phillips},
  {Richards}, {Toledo}, {Aladro}, {Broguiere}, {Cortes}, {Cortes}, {Espada},
  {Galarza}, {Garcia-Appadoo}, {Guzman-Ramirez}, {Humphreys}, {Jung}, {Kameno},
  {Laing}, {Leon}, {Marconi}, {Mignano}, {Nikolic}, {Nyman}, {Radiszcz},
  {Remijan}, {Rod{\'o}n}, {Sawada}, {Takahashi}, {Tilanus}, {Vila Vilaro},
  {Watson}, {Wiklind}, {Akiyama}, {Chapillon}, {de Gregorio-Monsalvo}, {Di
  Francesco}, {Gueth}, {Kawamura}, {Lee}, {Nguyen Luong}, {Mangum}, {Pietu},
  {Sanhueza}, {Saigo}, {Takakuwa}, {Ubach}, {van Kempen}, {Wootten},
  {Castro-Carrizo}, {Francke}, {Gallardo}, {Garcia}, {Gonzalez}, {Hill},
  {Kaminski}, {Kurono}, {Liu}, {Lopez}, {Morales}, {Plarre}, {Schieven},
  {Testi}, {Videla}, {Villard}, {Andreani}, {Hibbard}, \&
  {Tatematsu}}]{alma2015}
{ALMA Partnership}, {Brogan}, C.~L., {P{\'e}rez}, L.~M., {et~al.} 2015, \apjl,
  808, L3

\bibitem[{{Andrews} {et~al.}(2011){Andrews}, {Wilner}, {Espaillat}, {Hughes},
  {Dullemond}, {McClure}, {Qi}, \& {Brown}}]{andrews2011}
{Andrews}, S.~M., {Wilner}, D.~J., {Espaillat}, C., {et~al.} 2011, \apj, 732,
  42

\bibitem[{{Andrews} {et~al.}(2009){Andrews}, {Wilner}, {Hughes}, {Qi}, \&
  {Dullemond}}]{andrews2009}
{Andrews}, S.~M., {Wilner}, D.~J., {Hughes}, A.~M., {Qi}, C., \& {Dullemond},
  C.~P. 2009, \apj, 700, 1502

\bibitem[{{Andrews} {et~al.}(2016){Andrews}, {Wilner}, {Zhu}, {Birnstiel},
  {Carpenter}, {P{\'e}rez}, {Bai}, {{\"O}berg}, {Hughes}, {Isella}, \&
  {Ricci}}]{andrews2016}
{Andrews}, S.~M., {Wilner}, D.~J., {Zhu}, Z., {et~al.} 2016, \apjl, 820, L40

\bibitem[{{Ansdell} {et~al.}(2017){Ansdell}, {Williams}, {Manara}, {Miotello},
  {Facchini}, {van der Marel}, {Testi}, \& {van Dishoeck}}]{ansdell2017}
{Ansdell}, M., {Williams}, J.~P., {Manara}, C.~F., {et~al.} 2017, \aj, 153, 240

\bibitem[{{Ansdell} {et~al.}(2016){Ansdell}, {Williams}, {van der Marel},
  {Carpenter}, {Guidi}, {Hogerheijde}, {Mathews}, {Manara}, {Miotello},
  {Natta}, {Oliveira}, {Tazzari}, {Testi}, {van Dishoeck}, \& {van
  Terwisga}}]{ansdell16a}
{Ansdell}, M., {Williams}, J.~P., {van der Marel}, N., {et~al.} 2016, \apj,
  828, 46

\bibitem[{{Ataiee} {et~al.}(2018){Ataiee}, {Baruteau}, {Alibert}, \&
  {Benz}}]{ataiee2018}
{Ataiee}, S., {Baruteau}, C., {Alibert}, Y., \& {Benz}, W. 2018, \aap, 615,
  A110

\bibitem[{{Balbus} \& {Hawley}(1991)}]{balbus1991}
{Balbus}, S.~A. \& {Hawley}, J.~F. 1991, \apj, 376, 214

\bibitem[{{Balbus} \& {Hawley}(1998)}]{balbus1998}
{Balbus}, S.~A. \& {Hawley}, J.~F. 1998, Reviews of Modern Physics, 70, 1

\bibitem[{{Ballabio} {et~al.}(2018){Ballabio}, {Dipierro}, {Veronesi},
  {Lodato}, {Hutchison}, {Laibe}, \& {Price}}]{ballabio18a}
{Ballabio}, G., {Dipierro}, G., {Veronesi}, B., {et~al.} 2018, \mnras, 477,
  2766

\bibitem[{{Banzatti} {et~al.}(2015){Banzatti}, {Pinilla}, {Ricci},
  {Pontoppidan}, {Birnstiel}, \& {Ciesla}}]{banzatti2015}
{Banzatti}, A., {Pinilla}, P., {Ricci}, L., {et~al.} 2015, \apjl, 815, L15

\bibitem[{{Bate} {et~al.}(1995){Bate}, {Bonnell}, \& {Price}}]{bate1995}
{Bate}, M.~R., {Bonnell}, I.~A., \& {Price}, N.~M. 1995, \mnras, 277, 362

\bibitem[{{Beichman} {et~al.}(1988){Beichman}, {Neugebauer}, {Habing}, {Clegg},
  \& {Chester}}]{beichman1988}
{Beichman}, C.~A., {Neugebauer}, G., {Habing}, H.~J., {Clegg}, P.~E., \&
  {Chester}, T.~J., eds. 1988, {Infrared astronomical satellite (IRAS) catalogs
  and atlases. Volume 1: Explanatory supplement}, Vol.~1

\bibitem[{{Bertrang} {et~al.}(2017){Bertrang}, {Flock}, \&
  {Wolf}}]{bertrang2017}
{Bertrang}, G.~H.-M., {Flock}, M., \& {Wolf}, S. 2017, \mnras, 464, L61

\bibitem[{{B{\'e}thune} {et~al.}(2016){B{\'e}thune}, {Lesur}, \&
  {Ferreira}}]{bethune2016}
{B{\'e}thune}, W., {Lesur}, G., \& {Ferreira}, J. 2016, \aap, 589, A87

\bibitem[{{Birnstiel} \& {Andrews}(2014)}]{birnstiel14a}
{Birnstiel}, T. \& {Andrews}, S.~M. 2014, \apj, 780, 153

\bibitem[{{Birnstiel} {et~al.}(2016){Birnstiel}, {Fang}, \&
  {Johansen}}]{birnstiel2016}
{Birnstiel}, T., {Fang}, M., \& {Johansen}, A. 2016, \ssr, 205, 41

\bibitem[{{Boehler} {et~al.}(2018){Boehler}, {Ricci}, {Weaver}, {Isella},
  {Benisty}, {Carpenter}, {Grady}, {Shen}, {Tang}, \& {Perez}}]{boehler2018}
{Boehler}, Y., {Ricci}, L., {Weaver}, E., {et~al.} 2018, \apj, 853, 162

\bibitem[{{Calvet} {et~al.}(1991){Calvet}, {Patino}, {Magris}, \&
  {D'Alessio}}]{calvet1991}
{Calvet}, N., {Patino}, A., {Magris}, G.~C., \& {D'Alessio}, P. 1991, \apj,
  380, 617

\bibitem[{{Casassus} {et~al.}(2013){Casassus}, {van der Plas}, {M}, {Dent},
  {Fomalont}, {Hagelberg}, {Hales}, {Jord{\'a}n}, {Mawet}, {M{\'e}nard},
  {Wootten}, {Wilner}, {Hughes}, {Schreiber}, {Girard}, {Ercolano}, {Canovas},
  {Rom{\'a}n}, \& {Salinas}}]{casassus2013}
{Casassus}, S., {van der Plas}, G., {M}, S.~P., {et~al.} 2013, \nat, 493, 191

\bibitem[{{Chiang} \& {Goldreich}(1997)}]{chiang1997}
{Chiang}, E.~I. \& {Goldreich}, P. 1997, \apj, 490, 368

\bibitem[{{Cieza} {et~al.}(2017){Cieza}, {Casassus}, {P{\'e}rez}, {Hales},
  {C{\'a}rcamo}, {Ansdell}, {Avenhaus}, {Bayo}, {Bertrang}, {C{\'a}novas},
  {Christiaens}, {Dent}, {Ferrero}, {Gamen}, {Olofsson}, {Orcajo}, {Osses},
  {Pe{\~n}a-Ramirez}, {Principe}, {Ru{\'{\i}}z-Rodr{\'{\i}}guez}, {Schreiber},
  {van der Plas}, {Williams}, \& {Zurlo}}]{cieza2017}
{Cieza}, L.~A., {Casassus}, S., {P{\'e}rez}, S., {et~al.} 2017, \apjl, 851, L23

\bibitem[{{Cieza} {et~al.}(2010){Cieza}, {Schreiber}, {Romero}, {Mora},
  {Merin}, {Swift}, {Orellana}, {Williams}, {Harvey}, \& {Evans}}]{cieza2010}
{Cieza}, L.~A., {Schreiber}, M.~R., {Romero}, G.~A., {et~al.} 2010, \apj, 712,
  925

\bibitem[{{Cossins} {et~al.}(2009){Cossins}, {Lodato}, \&
  {Clarke}}]{cossins09a}
{Cossins}, P., {Lodato}, G., \& {Clarke}, C.~J. 2009, \mnras, 393, 1157

\bibitem[{{Crida} {et~al.}(2006){Crida}, {Morbidelli}, \& {Masset}}]{crida2006}
{Crida}, A., {Morbidelli}, A., \& {Masset}, F. 2006, \icarus, 181, 587

\bibitem[{{Cutri} \& {et al.}(2013)}]{cutri2013}
{Cutri}, R.~M. \& {et al.} 2013, VizieR Online Data Catalog, 2328

\bibitem[{{Cutri} {et~al.}(2003){Cutri}, {Skrutskie}, {van Dyk}, {Beichman},
  {Carpenter}, {Chester}, {Cambresy}, {Evans}, {Fowler}, {Gizis}, {Howard},
  {Huchra}, {Jarrett}, {Kopan}, {Kirkpatrick}, {Light}, {Marsh}, {McCallon},
  {Schneider}, {Stiening}, {Sykes}, {Weinberg}, {Wheaton}, {Wheelock}, \&
  {Zacarias}}]{cutri2003}
{Cutri}, R.~M., {Skrutskie}, M.~F., {van Dyk}, S., {et~al.} 2003, VizieR Online
  Data Catalog, 2246

\bibitem[{{D'Alessio} {et~al.}(2001){D'Alessio}, {Calvet}, \&
  {Hartmann}}]{dalessio2001}
{D'Alessio}, P., {Calvet}, N., \& {Hartmann}, L. 2001, \apj, 553, 321

\bibitem[{{D'Angelo} {et~al.}(2010){D'Angelo}, {Durisen}, \&
  {Lissauer}}]{dangelo10a}
{D'Angelo}, G., {Durisen}, R.~H., \& {Lissauer}, J.~J. 2010, {Giant Planet
  Formation}, ed. S.~{Seager}, 319--346

\bibitem[{{D'Angelo} \& {Lubow}(2008)}]{dangelo08a}
{D'Angelo}, G. \& {Lubow}, S.~H. 2008, \apj, 685, 560

\bibitem[{{Dipierro} \& {Laibe}(2017)}]{dipierro2017}
{Dipierro}, G. \& {Laibe}, G. 2017, \mnras, 469, 1932

\bibitem[{{Dipierro} {et~al.}(2015{\natexlab{a}}){Dipierro}, {Pinilla},
  {Lodato}, \& {Testi}}]{dipierro15a}
{Dipierro}, G., {Pinilla}, P., {Lodato}, G., \& {Testi}, L. 2015{\natexlab{a}},
  \mnras, 451, 974

\bibitem[{{Dipierro} {et~al.}(2015{\natexlab{b}}){Dipierro}, {Price}, {Laibe},
  {Hirsh}, {Cerioli}, \& {Lodato}}]{dipierro2015}
{Dipierro}, G., {Price}, D., {Laibe}, G., {et~al.} 2015{\natexlab{b}}, \mnras,
  453, L73

\bibitem[{{Dipierro} {et~al.}(2018){Dipierro}, {Ricci}, {P{\'e}rez}, {Lodato},
  {Alexander}, {Laibe}, {Andrews}, {Carpenter}, {Chandler}, {Greaves}, {Hall},
  {Henning}, {Kwon}, {Linz}, {Mundy}, {Sargent}, {Tazzari}, {Testi}, \&
  {Wilner}}]{dipierro2018}
{Dipierro}, G., {Ricci}, L., {P{\'e}rez}, L., {et~al.} 2018, \mnras, 475, 5296

\bibitem[{{Donehew} \& {Brittain}(2011)}]{donehew2011}
{Donehew}, B. \& {Brittain}, S. 2011, \aj, 141, 46

\bibitem[{{Dong} {et~al.}(2018){Dong}, {Liu}, {Eisner}, {Andrews}, {Fung},
  {Zhu}, {Chiang}, {Hashimoto}, {Liu}, {Casassus}, {Esposito}, {Hasegawa},
  {Muto}, {Pavlyuchenkov}, {Wilner}, {Akiyama}, {Tamura}, \&
  {Wisniewski}}]{dong2018}
{Dong}, R., {Liu}, S.-y., {Eisner}, J., {et~al.} 2018, \apj, 860, 124

\bibitem[{{Dorschner} {et~al.}(1995){Dorschner}, {Begemann}, {Henning},
  {Jaeger}, \& {Mutschke}}]{dorschner1995}
{Dorschner}, J., {Begemann}, B., {Henning}, T., {Jaeger}, C., \& {Mutschke}, H.
  1995, \aap, 300, 503

\bibitem[{{Dubrulle} {et~al.}(1995){Dubrulle}, {Morfill}, \&
  {Sterzik}}]{dubrulle95a}
{Dubrulle}, B., {Morfill}, G., \& {Sterzik}, M. 1995, \icarus, 114, 237

\bibitem[{{Dullemond} \& {Dominik}(2004)}]{dullemond2004}
{Dullemond}, C.~P. \& {Dominik}, C. 2004, \aap, 421, 1075

\bibitem[{{Dullemond} {et~al.}(2001){Dullemond}, {Dominik}, \&
  {Natta}}]{dullemond2001}
{Dullemond}, C.~P., {Dominik}, C., \& {Natta}, A. 2001, \apj, 560, 957

\bibitem[{{Dullemond} {et~al.}(2012){Dullemond}, {Juhasz}, {Pohl}, {Sereshti},
  {Shetty}, {Peters}, {Commercon}, \& {Flock}}]{radmc3d2012}
{Dullemond}, C.~P., {Juhasz}, A., {Pohl}, A., {et~al.} 2012, {RADMC-3D: A
  multi-purpose radiative transfer tool}, Astrophysics Source Code Library

\bibitem[{{Espaillat} {et~al.}(2007){Espaillat}, {Calvet}, {D'Alessio},
  {Hern{\'a}ndez}, {Qi}, {Hartmann}, {Furlan}, \& {Watson}}]{espaillat2007}
{Espaillat}, C., {Calvet}, N., {D'Alessio}, P., {et~al.} 2007, \apjl, 670, L135

\bibitem[{{Espaillat} {et~al.}(2010){Espaillat}, {D'Alessio}, {Hern{\'a}ndez},
  {Nagel}, {Luhman}, {Watson}, {Calvet}, {Muzerolle}, \&
  {McClure}}]{espaillat2010}
{Espaillat}, C., {D'Alessio}, P., {Hern{\'a}ndez}, J., {et~al.} 2010, \apj,
  717, 441

\bibitem[{{Espaillat} {et~al.}(2014){Espaillat}, {Muzerolle}, {Najita},
  {Andrews}, {Zhu}, {Calvet}, {Kraus}, {Hashimoto}, {Kraus}, \&
  {D'Alessio}}]{espaillat2014}
{Espaillat}, C., {Muzerolle}, J., {Najita}, J., {et~al.} 2014, Protostars and
  Planets VI, 497

\bibitem[{{Facchini} {et~al.}(2018){Facchini}, {Pinilla}, {van Dishoeck}, \&
  {de Juan Ovelar}}]{facchini2018}
{Facchini}, S., {Pinilla}, P., {van Dishoeck}, E.~F., \& {de Juan Ovelar}, M.
  2018, \aap, 612, A104

\bibitem[{{Fang} {et~al.}(2017){Fang}, {Sicilia-Aguilar}, {Wilner}, {Wang},
  {Roccatagliata}, {Fedele}, \& {Wang}}]{fang2017}
{Fang}, M., {Sicilia-Aguilar}, A., {Wilner}, D., {et~al.} 2017, \aap, 603, A132

\bibitem[{{Fang} {et~al.}(2009){Fang}, {van Boekel}, {Wang}, {Carmona},
  {Sicilia-Aguilar}, \& {Henning}}]{fang2009}
{Fang}, M., {van Boekel}, R., {Wang}, W., {et~al.} 2009, \aap, 504, 461

\bibitem[{{Fedele} {et~al.}(2018){Fedele}, {Tazzari}, {Booth}, {Testi},
  {Clarke}, {Pascucci}, {Kospal}, {Semenov}, {Bruderer}, {Henning}, \&
  {Teague}}]{fedele2018}
{Fedele}, D., {Tazzari}, M., {Booth}, R., {et~al.} 2018, \aap, 610, A24

\bibitem[{{Fernandes} {et~al.}(2018){Fernandes}, {Long}, {Pikhartova}, {Sitko},
  {Grady}, {Russell}, {Luria}, {Tyler}, {Bayyari}, {Danchi}, \&
  {Wisniewski}}]{fernandes2018}
{Fernandes}, R.~B., {Long}, Z.~C., {Pikhartova}, M., {et~al.} 2018, \apj, 856,
  103

\bibitem[{{Flaherty} {et~al.}(2015){Flaherty}, {Hughes}, {Rosenfeld},
  {Andrews}, {Chiang}, {Simon}, {Kerzner}, \& {Wilner}}]{flaherty2015}
{Flaherty}, K.~M., {Hughes}, A.~M., {Rosenfeld}, K.~A., {et~al.} 2015, \apj,
  813, 99

\bibitem[{{Flock} {et~al.}(2017){Flock}, {Nelson}, {Turner}, {Bertrang},
  {Carrasco-Gonz{\'a}lez}, {Henning}, {Lyra}, \& {Teague}}]{flock2017}
{Flock}, M., {Nelson}, R.~P., {Turner}, N.~J., {et~al.} 2017, \apj, 850, 131

\bibitem[{{Flock} {et~al.}(2015){Flock}, {Ruge}, {Dzyurkevich}, {Henning},
  {Klahr}, \& {Wolf}}]{flock2015}
{Flock}, M., {Ruge}, J.~P., {Dzyurkevich}, N., {et~al.} 2015, \aap, 574, A68

\bibitem[{{Fouchet} {et~al.}(2010){Fouchet}, {Gonzalez}, \&
  {Maddison}}]{fouchet10a}
{Fouchet}, L., {Gonzalez}, J.-F., \& {Maddison}, S.~T. 2010, \aap, 518, A16

\bibitem[{{Fouchet} {et~al.}(2007){Fouchet}, {Maddison}, {Gonzalez}, \&
  {Murray}}]{fouchet07a}
{Fouchet}, L., {Maddison}, S.~T., {Gonzalez}, J.-F., \& {Murray}, J.~R. 2007,
  \aap, 474, 1037

\bibitem[{{Fromang} \& {Nelson}(2009)}]{fromang09a}
{Fromang}, S. \& {Nelson}, R.~P. 2009, \aap, 496, 597

\bibitem[{{Gaia Collaboration} {et~al.}(2018){Gaia Collaboration}, {Brown},
  {Vallenari}, {Prusti}, {de Bruijne}, {Babusiaux}, \&
  {Bailer-Jones}}]{gaia2018}
{Gaia Collaboration}, {Brown}, A.~G.~A., {Vallenari}, A., {et~al.} 2018, ArXiv
  e-prints

\bibitem[{{Gaia Collaboration} {et~al.}(2016){Gaia Collaboration}, {Prusti},
  {de Bruijne}, {Brown}, {Vallenari}, {Babusiaux}, {Bailer-Jones}, {Bastian},
  {Biermann}, {Evans}, \& et~al.}]{gaia2016}
{Gaia Collaboration}, {Prusti}, T., {de Bruijne}, J.~H.~J., {et~al.} 2016,
  \aap, 595, A1

\bibitem[{{Garufi} {et~al.}(2014){Garufi}, {Podio}, {Kamp}, {M{\'e}nard},
  {Brittain}, {Eiroa}, {Montesinos}, {Alonso-Mart{\'{\i}}nez}, {Thi}, \&
  {Woitke}}]{Garufi2014}
{Garufi}, A., {Podio}, L., {Kamp}, I., {et~al.} 2014, \aap, 567, A141

\bibitem[{{Gr{\"a}fe} {et~al.}(2013){Gr{\"a}fe}, {Wolf}, {Guilloteau},
  {Dutrey}, {Stapelfeldt}, {Pontoppidan}, \& {Sauter}}]{grafe2013}
{Gr{\"a}fe}, C., {Wolf}, S., {Guilloteau}, S., {et~al.} 2013, \aap, 553, A69

\bibitem[{{Guilloteau} {et~al.}(2011){Guilloteau}, {Dutrey}, {Pi{\'e}tu}, \&
  {Boehler}}]{guilloteau2011}
{Guilloteau}, S., {Dutrey}, A., {Pi{\'e}tu}, V., \& {Boehler}, Y. 2011, \aap,
  529, A105

\bibitem[{{Hamidouche} {et~al.}(2006){Hamidouche}, {Looney}, \&
  {Mundy}}]{hamidouche2006}
{Hamidouche}, M., {Looney}, L.~W., \& {Mundy}, L.~G. 2006, \apj, 651, 321

\bibitem[{{Hendler} {et~al.}(2017){Hendler}, {Pinilla}, {Pascucci}, {Pohl},
  {Mulders}, {Henning}, {Dong}, {Clarke}, {Owen}, \&
  {Hollenbach}}]{hendler2017}
{Hendler}, N.~P., {Pinilla}, P., {Pascucci}, I., {et~al.} 2017, ArXiv e-prints

\bibitem[{{Huang} {et~al.}(2017){Huang}, {{\"O}berg}, {Qi}, {Aikawa},
  {Andrews}, {Furuya}, {Guzm{\'a}n}, {Loomis}, {van Dishoeck}, \&
  {Wilner}}]{huang2017}
{Huang}, J., {{\"O}berg}, K.~I., {Qi}, C., {et~al.} 2017, \apj, 835, 231

\bibitem[{{Hughes} {et~al.}(2008){Hughes}, {Wilner}, {Qi}, \&
  {Hogerheijde}}]{hughes2008}
{Hughes}, A.~M., {Wilner}, D.~J., {Qi}, C., \& {Hogerheijde}, M.~R. 2008, \apj,
  678, 1119

\bibitem[{{Isella} {et~al.}(2010{\natexlab{a}}){Isella}, {Carpenter}, \&
  {Sargent}}]{isella2010a}
{Isella}, A., {Carpenter}, J.~M., \& {Sargent}, A.~I. 2010{\natexlab{a}}, \apj,
  714, 1746

\bibitem[{{Isella} {et~al.}(2016){Isella}, {Guidi}, {Testi}, {Liu}, {Li}, \&
  et~al.}]{isella2016}
{Isella}, A., {Guidi}, G., {Testi}, L., {et~al.} 2016, PhRvL, 117, 251101

\bibitem[{{Isella} {et~al.}(2010{\natexlab{b}}){Isella}, {Natta}, {Wilner},
  {Carpenter}, \& {Testi}}]{isella2010b}
{Isella}, A., {Natta}, A., {Wilner}, D., {Carpenter}, J.~M., \& {Testi}, L.
  2010{\natexlab{b}}, \apj, 725, 1735

\bibitem[{{Ishihara} {et~al.}(2010){Ishihara}, {Onaka}, {Kataza}, {Salama},
  {Alfageme}, {Cassatella}, {Cox}, {Garc{\'\i}a-Lario}, {Stephenson}, {Cohen},
  {Fujishiro}, {Fujiwara}, {Hasegawa}, {Ita}, {Kim}, {Matsuhara}, {Murakami},
  {M{\"u}ller}, {Nakagawa}, {Ohyama}, {Oyabu}, {Pyo}, {Sakon}, {Shibai},
  {Takita}, {Tanab{\'e}}, {Uemizu}, {Ueno}, {Usui}, {Wada}, {Watarai},
  {Yamamura}, \& {Yamauchi}}]{ishihara2010}
{Ishihara}, D., {Onaka}, T., {Kataza}, H., {et~al.} 2010, \aap, 514, A1

\bibitem[{{J\"ager} {et~al.}(1998){J\"ager}, {Mutschke}, \&
  {Henning}}]{jager1998}
{J\"ager}, C., {Mutschke}, H., \& {Henning}, T. 1998, \aap, 332, 291

\bibitem[{{Jin} {et~al.}(2016){Jin}, {Li}, {Isella}, {Li}, \& {Ji}}]{jin2016}
{Jin}, S., {Li}, S., {Isella}, A., {Li}, H., \& {Ji}, J. 2016, \apj, 818, 76

\bibitem[{{Juh{\'a}sz} {et~al.}(2010){Juh{\'a}sz}, {Bouwman}, {Henning},
  {Acke}, {van den Ancker}, {Meeus}, {Dominik}, {Min}, {Tielens}, \&
  {Waters}}]{juhasz2010}
{Juh{\'a}sz}, A., {Bouwman}, J., {Henning}, T., {et~al.} 2010, \apj, 721, 431

\bibitem[{{Kenyon} \& {Hartmann}(1995)}]{kenyon1995}
{Kenyon}, S.~J. \& {Hartmann}, L. 1995, \apjs, 101, 117

\bibitem[{{Keppler} {et~al.}(2018){Keppler}, {Benisty}, {M{\"u}ller},
  {Henning}, {van Boekel}, {Cantalloube}, {Ginski}, {van Holstein}, {Maire},
  {Pohl}, {Samland}, {Avenhaus}, {Baudino}, {Boccaletti}, {de Boer},
  {Bonnefoy}, {Chauvin}, {Desidera}, {Langlois}, {Lazzoni}, {Marleau},
  {Mordasini}, {Pawellek}, {Stolker}, {Vigan}, {Zurlo}, {Birnstiel},
  {Brandner}, {Feldt}, {Flock}, {Girard}, {Gratton}, {Hagelberg}, {Isella},
  {Janson}, {Juhasz}, {Kemmer}, {Kral}, {Lagrange}, {Launhardt}, {Matter},
  {M{\'e}nard}, {Milli}, {Molli{\`e}re}, {Olofsson}, {Perez}, {Pinilla},
  {Pinte}, {Quanz}, {Schmidt}, {Udry}, {Wahhaj}, {Williams}, {Buenzli},
  {Cudel}, {Dominik}, {Galicher}, {Kasper}, {Lannier}, {Mesa}, {Mouillet},
  {Peretti}, {Perrot}, {Salter}, {Sissa}, {Wildi}, {Abe}, {Antichi},
  {Augereau}, {Baruffolo}, {Baudoz}, {Bazzon}, {Beuzit}, {Blanchard}, {Brems},
  {Buey}, {De Caprio}, {Carbillet}, {Carle}, {Cascone}, {Cheetham}, {Claudi},
  {Costille}, {Delboulb{\'e}}, {Dohlen}, {Fantinel}, {Feautrier}, {Fusco},
  {Giro}, {Gisler}, {Gluck}, {Gry}, {Hubin}, {Hugot}, {Jaquet}, {Le Mignant},
  {Llored}, {Madec}, {Magnard}, {Martinez}, {Maurel}, {Meyer},
  {Moeller-Nilsson}, {Moulin}, {Mugnier}, {Origne}, {Pavlov}, {Perret},
  {Petit}, {Pragt}, {Puget}, {Rabou}, {Ramos}, {Rigal}, {Rochat}, {Roelfsema},
  {Rousset}, {Roux}, {Salasnich}, {Sauvage}, {Sevin}, {Soenke}, {Stadler},
  {Suarez}, {Turatto}, \& {Weber}}]{keppler2018}
{Keppler}, M., {Benisty}, M., {M{\"u}ller}, A., {et~al.} 2018, ArXiv e-prints

\bibitem[{{Kirchschlager} {et~al.}(2016){Kirchschlager}, {Wolf}, \&
  {Madlener}}]{Kirchschlager2016}
{Kirchschlager}, F., {Wolf}, S., \& {Madlener}, D. 2016, \mnras, 462, 858

\bibitem[{{Kirkpatrick} {et~al.}(1983){Kirkpatrick}, {Gelatt}, \&
  {Vecchi}}]{kirkpatrick1983}
{Kirkpatrick}, S., {Gelatt}, C.~D., \& {Vecchi}, M.~P. 1983, Science, 220, 671

\bibitem[{{Kratter} \& {Lodato}(2016)}]{kratter16a}
{Kratter}, K.~M. \& {Lodato}, G. 2016, ArXiv e-prints

\bibitem[{{Kraus} \& {Ireland}(2012)}]{kraus2012}
{Kraus}, A.~L. \& {Ireland}, M.~J. 2012, \apj, 745, 5

\bibitem[{{Kraus} {et~al.}(2017){Kraus}, {Kreplin}, {Fukugawa}, {Muto},
  {Sitko}, {Young}, {Bate}, {Grady}, {Harries}, {Monnier}, {Willson}, \&
  {Wisniewski}}]{kraus2017}
{Kraus}, S., {Kreplin}, A., {Fukugawa}, M., {et~al.} 2017, \apjl, 848, L11

\bibitem[{{Kurucz}(1994)}]{Kurucz1994}
{Kurucz}, R. 1994, Solar abundance model atmospheres for 0,1,2,4,8 km/s.~Kurucz
  CD-ROM No.~19.~ Cambridge, Mass.: Smithsonian Astrophysical Observatory,
  1994., 19

\bibitem[{{Kusakabe} {et~al.}(2012){Kusakabe}, {Grady}, {Sitko}, {Hashimoto},
  {Kudo}, {Fukagawa}, {Muto}, {Wisniewski}, {Min}, {Mayama}, {Werren}, {Day},
  {Beerman}, {Lynch}, {Russell}, {Brafford}, {Kuzuhara}, {Brandt}, {Abe},
  {Brandner}, {Carson}, {Egner}, {Feldt}, {Goto}, {Guyon}, {Hayano}, {Hayashi},
  {Hayashi}, {Henning}, {Hodapp}, {Ishii}, {Iye}, {Janson}, {Kandori}, {Knapp},
  {Matsuo}, {McElwain}, {Miyama}, {Morino}, {Moro-Martin}, {Nishimura}, {Pyo},
  {Suto}, {Suzuki}, {Takami}, {Takato}, {Terada}, {Thalmann}, {Tomono},
  {Turner}, {Watanabe}, {Yamada}, {Takami}, {Usuda}, \&
  {Tamura}}]{kusakabe2012}
{Kusakabe}, N., {Grady}, C.~A., {Sitko}, M.~L., {et~al.} 2012, \apj, 753, 153

\bibitem[{{Liu} {et~al.}(2017){Liu}, {Henning}, {Carrasco-Gonz{\'a}lez},
  {Chandler}, {Linz}, {Birnstiel}, {van Boekel}, {P{\'e}rez}, {Flock}, {Testi},
  {Rodr{\'{\i}}guez}, \& {Galv{\'a}n-Madrid}}]{liu2017}
{Liu}, Y., {Henning}, T., {Carrasco-Gonz{\'a}lez}, C., {et~al.} 2017, \aap,
  607, A74

\bibitem[{{Liu} {et~al.}(2012){Liu}, {Madlener}, {Wolf}, {Wang}, \&
  {Ruge}}]{lium2012}
{Liu}, Y., {Madlener}, D., {Wolf}, S., {Wang}, H., \& {Ruge}, J.~P. 2012, \aap,
  546, A7

\bibitem[{{Lodato} \& {Price}(2010)}]{lodato10a}
{Lodato}, G. \& {Price}, D.~J. 2010, \mnras, 405, 1212

\bibitem[{{Long} {et~al.}(2017){Long}, {Herczeg}, {Pascucci}, {Drabek-Maunder},
  {Mohanty}, {Testi}, {Apai}, {Hendler}, {Henning}, {Manara}, \&
  {Mulders}}]{long2017}
{Long}, F., {Herczeg}, G.~J., {Pascucci}, I., {et~al.} 2017, \apj, 844, 99

\bibitem[{{Long} {et~al.}(2018){Long}, {Pinilla}, {Herczeg}, {Harsono},
  {Dipierro}, {Pascucci}, {Hendler}, {Tazzari}, {Ragusa}, {Salyk}, {Edwards},
  {Lodato}, {van de Plas}, {Johnstone}, {Liu}, {Boehler}, {Cabrit}, {Manara},
  {Menard}, {Mulders}, {Nisini}, {Fischer}, {Rigliaco}, {Banzatti}, {Avenhaus},
  \& {Gully-Santiago}}]{long2018}
{Long}, F., {Pinilla}, P., {Herczeg}, G.~J., {et~al.} 2018, ArXiv e-prints

\bibitem[{{Louvet} {et~al.}(2018){Louvet}, {Dougados}, {Cabrit}, {Mardones},
  {M{\'e}nard}, {Tabone}, {Pinte}, \& {Dent}}]{louvet2018}
{Louvet}, F., {Dougados}, C., {Cabrit}, S., {et~al.} 2018, ArXiv e-prints

\bibitem[{{Luhman} {et~al.}(2017){Luhman}, {Mamajek}, {Shukla}, \&
  {Loutrel}}]{luhman2017}
{Luhman}, K.~L., {Mamajek}, E.~E., {Shukla}, S.~J., \& {Loutrel}, N.~P. 2017,
  \aj, 153, 46

\bibitem[{{Madlener} {et~al.}(2012){Madlener}, {Wolf}, {Dutrey}, \&
  {Guilloteau}}]{madlener2012}
{Madlener}, D., {Wolf}, S., {Dutrey}, A., \& {Guilloteau}, S. 2012, \aap, 543,
  A81

\bibitem[{{Maire} {et~al.}(2017){Maire}, {Stolker}, {Messina}, {M{\"u}ller},
  {Biller}, {Currie}, {Dominik}, {Grady}, {Boccaletti}, {Bonnefoy}, {Chauvin},
  {Galicher}, {Millward}, {Pohl}, {Brandner}, {Henning}, {Lagrange},
  {Langlois}, {Meyer}, {Quanz}, {Vigan}, {Zurlo}, {van Boekel}, {Buenzli},
  {Buey}, {Desidera}, {Feldt}, {Fusco}, {Ginski}, {Giro}, {Gratton}, {Hubin},
  {Lannier}, {Le Mignant}, {Mesa}, {Peretti}, {Perrot}, {Ramos}, {Salter},
  {Samland}, {Sissa}, {Stadler}, {Thalmann}, {Udry}, \& {Weber}}]{maire2017}
{Maire}, A.-L., {Stolker}, T., {Messina}, S., {et~al.} 2017, \aap, 601, A134

\bibitem[{{Mannings} \& {Sargent}(1997)}]{mannings1997}
{Mannings}, V. \& {Sargent}, A.~I. 1997, \apj, 490, 792

\bibitem[{{McMullin} {et~al.}(2007){McMullin}, {Waters}, {Schiebel}, {Young},
  \& {Golap}}]{mcmullin07a}
{McMullin}, J.~P., {Waters}, B., {Schiebel}, D., {Young}, W., \& {Golap}, K.
  2007, in Astronomical Society of the Pacific Conference Series, Vol. 376,
  Astronomical Data Analysis Software and Systems XVI, ed. R.~A. {Shaw},
  F.~{Hill}, \& D.~J. {Bell}, 127

\bibitem[{{Meeus} {et~al.}(2001){Meeus}, {Waters}, {Bouwman}, {van den Ancker},
  {Waelkens}, \& {Malfait}}]{meeus2001}
{Meeus}, G., {Waters}, L.~B.~F.~M., {Bouwman}, J., {et~al.} 2001, \aap, 365,
  476

\bibitem[{{Mendigut{\'{\i}}a} {et~al.}(2013){Mendigut{\'{\i}}a}, {Brittain},
  {Eiroa}, {Meeus}, {Montesinos}, {Mora}, {Muzerolle}, {Oudmaijer}, \&
  {Rigliaco}}]{mendigutia2013}
{Mendigut{\'{\i}}a}, I., {Brittain}, S., {Eiroa}, C., {et~al.} 2013, \apj, 776,
  44

\bibitem[{{Mendigut{\'{\i}}a} {et~al.}(2012){Mendigut{\'{\i}}a}, {Mora},
  {Montesinos}, {Eiroa}, {Meeus}, {Mer{\'{\i}}n}, \&
  {Oudmaijer}}]{mendigutia2012}
{Mendigut{\'{\i}}a}, I., {Mora}, A., {Montesinos}, B., {et~al.} 2012, \aap,
  543, A59

\bibitem[{{Mer{\'{\i}}n} {et~al.}(2010){Mer{\'{\i}}n}, {Brown}, {Oliveira},
  {Herczeg}, {van Dishoeck}, {Bottinelli}, {Evans}, {Cieza}, {Spezzi},
  {Alcal{\'a}}, {Harvey}, {Blake}, {Bayo}, {Geers}, {Lahuis}, {Prusti},
  {Augereau}, {Olofsson}, {Walter}, \& {Chiu}}]{merin2010}
{Mer{\'{\i}}n}, B., {Brown}, J.~M., {Oliveira}, I., {et~al.} 2010, \apj, 718,
  1200

\bibitem[{{Millan-Gabet} {et~al.}(2016){Millan-Gabet}, {Che}, {Monnier},
  {Sitko}, {Russell}, {Grady}, {Day}, {Perry}, {Harries}, {Aarnio}, {Colavita},
  {Wizinowich}, {Ragland}, \& {Woillez}}]{millan-gabet2016}
{Millan-Gabet}, R., {Che}, X., {Monnier}, J.~D., {et~al.} 2016, \apj, 826, 120

\bibitem[{{Miyake} \& {Nakagawa}(1995)}]{miyake1995}
{Miyake}, K. \& {Nakagawa}, Y. 1995, \apj, 441, 361

\bibitem[{{Mora} {et~al.}(2001){Mora}, {Mer{\'{\i}}n}, {Solano}, {Montesinos},
  {de Winter}, {Eiroa}, {Ferlet}, {Grady}, {Davies}, {Miranda}, {Oudmaijer},
  {Palacios}, {Quirrenbach}, {Harris}, {Rauer}, {Collier Cameron}, {Deeg},
  {Garz{\'o}n}, {Penny}, {Schneider}, {Tsapras}, \& {Wesselius}}]{mora2001}
{Mora}, A., {Mer{\'{\i}}n}, B., {Solano}, E., {et~al.} 2001, \aap, 378, 116

\bibitem[{{Muto} {et~al.}(2012){Muto}, {Grady}, {Hashimoto}, {Fukagawa},
  {Hornbeck}, {Sitko}, {Russell}, {Werren}, {Cur{\'e}}, {Currie}, {Ohashi},
  {Okamoto}, {Momose}, {Honda}, {Inutsuka}, {Takeuchi}, {Dong}, {Abe},
  {Brandner}, {Brandt}, {Carson}, {Egner}, {Feldt}, {Fukue}, {Goto}, {Guyon},
  {Hayano}, {Hayashi}, {Hayashi}, {Henning}, {Hodapp}, {Ishii}, {Iye},
  {Janson}, {Kandori}, {Knapp}, {Kudo}, {Kusakabe}, {Kuzuhara}, {Matsuo},
  {Mayama}, {McElwain}, {Miyama}, {Morino}, {Moro-Martin}, {Nishimura}, {Pyo},
  {Serabyn}, {Suto}, {Suzuki}, {Takami}, {Takato}, {Terada}, {Thalmann},
  {Tomono}, {Turner}, {Watanabe}, {Wisniewski}, {Yamada}, {Takami}, {Usuda}, \&
  {Tamura}}]{muto2012}
{Muto}, T., {Grady}, C.~A., {Hashimoto}, J., {et~al.} 2012, \apjl, 748, L22

\bibitem[{{{\"O}berg} {et~al.}(2011){{\"O}berg}, {Murray-Clay}, \&
  {Bergin}}]{oberg2011}
{{\"O}berg}, K.~I., {Murray-Clay}, R., \& {Bergin}, E.~A. 2011, \apjl, 743, L16

\bibitem[{{{\"O}berg} {et~al.}(2010){{\"O}berg}, {Qi}, {Fogel}, {Bergin},
  {Andrews}, {Espaillat}, {van Kempen}, {Wilner}, \& {Pascucci}}]{oberg2010}
{{\"O}berg}, K.~I., {Qi}, C., {Fogel}, J.~K.~J., {et~al.} 2010, \apj, 720, 480

\bibitem[{{Okuzumi} {et~al.}(2016){Okuzumi}, {Momose}, {Sirono}, {Kobayashi},
  \& {Tanaka}}]{okuzumi2016}
{Okuzumi}, S., {Momose}, M., {Sirono}, S.-i., {Kobayashi}, H., \& {Tanaka}, H.
  2016, \apj, 821, 82

\bibitem[{{Paardekooper} {et~al.}(2010){Paardekooper}, {Baruteau}, {Crida}, \&
  {Kley}}]{paardekooper2010}
{Paardekooper}, S.-J., {Baruteau}, C., {Crida}, A., \& {Kley}, W. 2010, \mnras,
  401, 1950

\bibitem[{{Paardekooper} \& {Mellema}(2006)}]{paardekooper06a}
{Paardekooper}, S.-J. \& {Mellema}, G. 2006, \aap, 453, 1129

\bibitem[{{Pascual} {et~al.}(2016){Pascual}, {Montesinos}, {Meeus}, {Marshall},
  {Mendigut{\'{\i}}a}, \& {Sandell}}]{pascual2016}
{Pascual}, N., {Montesinos}, B., {Meeus}, G., {et~al.} 2016, \aap, 586, A6

\bibitem[{{P{\'e}rez} {et~al.}(2016){P{\'e}rez}, {Carpenter}, {Andrews},
  {Ricci}, {Isella}, {Linz}, {Sargent}, {Wilner}, {Henning}, {Deller},
  {Chandler}, {Dullemond}, {Lazio}, {Menten}, {Corder}, {Storm}, {Testi},
  {Tazzari}, {Kwon}, {Calvet}, {Greaves}, {Harris}, \& {Mundy}}]{perez2016}
{P{\'e}rez}, L.~M., {Carpenter}, J.~M., {Andrews}, S.~M., {et~al.} 2016,
  Science, 353, 1519

\bibitem[{{Pi{\'e}tu} {et~al.}(2007){Pi{\'e}tu}, {Dutrey}, \&
  {Guilloteau}}]{pietu2007}
{Pi{\'e}tu}, V., {Dutrey}, A., \& {Guilloteau}, S. 2007, \aap, 467, 163

\bibitem[{{Pi{\'e}tu} {et~al.}(2006){Pi{\'e}tu}, {Dutrey}, {Guilloteau},
  {Chapillon}, \& {Pety}}]{pietu2006}
{Pi{\'e}tu}, V., {Dutrey}, A., {Guilloteau}, S., {Chapillon}, E., \& {Pety}, J.
  2006, \aap, 460, L43

\bibitem[{{Pinilla} {et~al.}(2015){Pinilla}, {de Juan Ovelar}, {Ataiee},
  {Benisty}, {Birnstiel}, {van Dishoeck}, \& {Min}}]{pinilla15c}
{Pinilla}, P., {de Juan Ovelar}, M., {Ataiee}, S., {et~al.} 2015, \aap, 573, A9

\bibitem[{{Pinilla} {et~al.}(2016){Pinilla}, {Flock}, {Ovelar}, \&
  {Birnstiel}}]{pinilla2016}
{Pinilla}, P., {Flock}, M., {Ovelar}, M.~d.~J., \& {Birnstiel}, T. 2016, \aap,
  596, A81

\bibitem[{{Pinilla} {et~al.}(2017){Pinilla}, {Pohl}, {Stammler}, \&
  {Birnstiel}}]{pinilla2017}
{Pinilla}, P., {Pohl}, A., {Stammler}, S.~M., \& {Birnstiel}, T. 2017, \apj,
  845, 68

\bibitem[{{Pinilla} {et~al.}(2018){Pinilla}, {Tazzari}, {Pascucci}, {Youdin},
  {Garufi}, {Manara}, {Testi}, {van der Plas}, {Barenfeld}, {Canovas}, {Cox},
  {Hendler}, {P{\'e}rez}, \& {van der Marel}}]{pinilla2018}
{Pinilla}, P., {Tazzari}, M., {Pascucci}, I., {et~al.} 2018, \apj, 859, 32

\bibitem[{{Pinte} {et~al.}(2016){Pinte}, {Dent}, {M{\'e}nard}, {Hales}, {Hill},
  {Cortes}, \& {de Gregorio-Monsalvo}}]{pinte2016}
{Pinte}, C., {Dent}, W.~R.~F., {M{\'e}nard}, F., {et~al.} 2016, \apj, 816, 25

\bibitem[{{Pinte} \& {Laibe}(2014)}]{pinte2014}
{Pinte}, C. \& {Laibe}, G. 2014, \aap, 565, A129

\bibitem[{{Price} \& {Laibe}(2015)}]{price15a}
{Price}, D.~J. \& {Laibe}, G. 2015, \mnras, 451, 813

\bibitem[{{Price} {et~al.}(2017){Price}, {Wurster}, {Nixon}, {Tricco},
  {Toupin}, {Pettitt}, {Chan}, {Laibe}, {Glover}, {Dobbs}, {Nealon}, {Liptai},
  {Worpel}, {Bonnerot}, {Dipierro}, {Ragusa}, {Federrath}, {Iaconi},
  {Reichardt}, {Forgan}, {Hutchison}, {Constantino}, {Ayliffe}, {Mentiplay},
  {Hirsh}, \& {Lodato}}]{price17a}
{Price}, D.~J., {Wurster}, J., {Nixon}, C., {et~al.} 2017, ArXiv e-prints

\bibitem[{{Rosotti} {et~al.}(2016){Rosotti}, {Juhasz}, {Booth}, \&
  {Clarke}}]{rosotti2016}
{Rosotti}, G.~P., {Juhasz}, A., {Booth}, R.~A., \& {Clarke}, C.~J. 2016,
  \mnras, 459, 2790

\bibitem[{{Ruane} {et~al.}(2017){Ruane}, {Mawet}, {Kastner}, {Meshkat},
  {Bottom}, {Femen{\'{\i}}a Castell{\'a}}, {Absil}, {Gomez Gonzalez}, {Huby},
  {Zhu}, {Jenson-Clem}, {Choquet}, \& {Serabyn}}]{ruane2017}
{Ruane}, G., {Mawet}, D., {Kastner}, J., {et~al.} 2017, \aj, 154, 73

\bibitem[{{Sallum} {et~al.}(2015){Sallum}, {Follette}, {Eisner}, {Close},
  {Hinz}, {Kratter}, {Males}, {Skemer}, {Macintosh}, {Tuthill}, {Bailey},
  {Defr{\`e}re}, {Morzinski}, {Rodigas}, {Spalding}, {Vaz}, \&
  {Weinberger}}]{sallum2015}
{Sallum}, S., {Follette}, K.~B., {Eisner}, J.~A., {et~al.} 2015, \nat, 527, 342

\bibitem[{{Salyk} {et~al.}(2013){Salyk}, {Herczeg}, {Brown}, {Blake},
  {Pontoppidan}, \& {van Dishoeck}}]{salyk2013}
{Salyk}, C., {Herczeg}, G.~J., {Brown}, J.~M., {et~al.} 2013, \apj, 769, 21

\bibitem[{{Sauter} {et~al.}(2009){Sauter}, {Wolf}, {Launhardt}, {Padgett},
  {Stapelfeldt}, {Pinte}, {Duch{\^e}ne}, {M{\'e}nard}, {McCabe}, {Pontoppidan},
  {Dunham}, {Bourke}, \& {Chen}}]{sauter2009}
{Sauter}, J., {Wolf}, S., {Launhardt}, R., {et~al.} 2009, \aap, 505, 1167

\bibitem[{{Shakura} \& {Sunyaev}(1973)}]{shakura73a}
{Shakura}, N.~I. \& {Sunyaev}, R.~A. 1973, \aap, 24, 337

\bibitem[{{Shu} {et~al.}(1987){Shu}, {Adams}, \& {Lizano}}]{shu1987}
{Shu}, F.~H., {Adams}, F.~C., \& {Lizano}, S. 1987, \araa, 25, 23

\bibitem[{{Simon} {et~al.}(2000){Simon}, {Dutrey}, \& {Guilloteau}}]{simon2000}
{Simon}, M., {Dutrey}, A., \& {Guilloteau}, S. 2000, \apj, 545, 1034

\bibitem[{{Sitko} {et~al.}(2008){Sitko}, {Carpenter}, {Kimes}, {Wilde},
  {Lynch}, {Russell}, {Rudy}, {Mazuk}, {Venturini}, {Puetter}, {Grady},
  {Polomski}, {Wisnewski}, {Brafford}, {Hammel}, \& {Perry}}]{sitko2008}
{Sitko}, M.~L., {Carpenter}, W.~J., {Kimes}, R.~L., {et~al.} 2008, \apj, 678,
  1070

\bibitem[{{Stammler} {et~al.}(2017){Stammler}, {Birnstiel}, {Pani{\'c}},
  {Dullemond}, \& {Dominik}}]{stammler2017}
{Stammler}, S.~M., {Birnstiel}, T., {Pani{\'c}}, O., {Dullemond}, C.~P., \&
  {Dominik}, C. 2017, \aap, 600, A140

\bibitem[{{Takahashi} \& {Inutsuka}(2016)}]{takahashi2016}
{Takahashi}, S.~Z. \& {Inutsuka}, S.-i. 2016, ArXiv e-prints

\bibitem[{{Teague} {et~al.}(2016){Teague}, {Guilloteau}, {Semenov}, {Henning},
  {Dutrey}, {Pi{\'e}tu}, {Birnstiel}, {Chapillon}, {Hollenbach}, \&
  {Gorti}}]{teague2016}
{Teague}, R., {Guilloteau}, S., {Semenov}, D., {et~al.} 2016, \aap, 592, A49

\bibitem[{{Terebey} {et~al.}(1984){Terebey}, {Shu}, \& {Cassen}}]{terebey1984}
{Terebey}, S., {Shu}, F.~H., \& {Cassen}, P. 1984, \apj, 286, 529

\bibitem[{{Testi} {et~al.}(2015){Testi}, {Skemer}, {Henning}, {Bailey},
  {Defr{\`e}re}, {Hinz}, {Leisenring}, {Vaz}, {Esposito}, {Fontana}, {Marconi},
  {Skrutskie}, \& {Veillet}}]{testi2015}
{Testi}, L., {Skemer}, A., {Henning}, T., {et~al.} 2015, \apjl, 812, L38

\bibitem[{{Tobin} {et~al.}(2016){Tobin}, {Kratter}, {Persson}, {Looney},
  {Dunham}, {Segura-Cox}, {Li}, {Chandler}, {Sadavoy}, {Harris}, {Melis}, \&
  {P{\'e}rez}}]{tobin16a}
{Tobin}, J.~J., {Kratter}, K.~M., {Persson}, M.~V., {et~al.} 2016, \nat, 538,
  483

\bibitem[{{Toomre}(1964)}]{toomre64a}
{Toomre}, A. 1964, \apj, 139, 1217

\bibitem[{{van Boekel} {et~al.}(2005){van Boekel}, {Min}, {Waters}, {de Koter},
  {Dominik}, {van den Ancker}, \& {Bouwman}}]{vanboekel2005}
{van Boekel}, R., {Min}, M., {Waters}, L.~B.~F.~M., {et~al.} 2005, \aap, 437,
  189

\bibitem[{{van der Marel} {et~al.}(2013){van der Marel}, {van Dishoeck},
  {Bruderer}, {Birnstiel}, {Pinilla}, {Dullemond}, {van Kempen}, {Schmalzl},
  {Brown}, {Herczeg}, {Mathews}, \& {Geers}}]{vanderMarel2013}
{van der Marel}, N., {van Dishoeck}, E.~F., {Bruderer}, S., {et~al.} 2013,
  Science, 340, 1199

\bibitem[{{van der Marel} {et~al.}(2016){van der Marel}, {Verhaar}, {van
  Terwisga}, {Mer{\'{\i}}n}, {Herczeg}, {Ligterink}, \& {van
  Dishoeck}}]{vandermarel2016}
{van der Marel}, N., {Verhaar}, B.~W., {van Terwisga}, S., {et~al.} 2016, \aap,
  592, A126

\bibitem[{{van der Plas} {et~al.}(2017){van der Plas}, {M{\'e}nard}, {Canovas},
  {Avenhaus}, {Casassus}, {Pinte}, {Caceres}, \& {Cieza}}]{vanderplas2017}
{van der Plas}, G., {M{\'e}nard}, F., {Canovas}, H., {et~al.} 2017, \aap, 607,
  A55

\bibitem[{{Weidenschilling}(1977)}]{weidenschilling77b}
{Weidenschilling}, S.~J. 1977, \apss, 51, 153

\bibitem[{{Williams} \& {Cieza}(2011)}]{williams2011}
{Williams}, J.~P. \& {Cieza}, L.~A. 2011, \araa, 49, 67

\bibitem[{{Wolf} {et~al.}(2003){Wolf}, {Padgett}, \& {Stapelfeldt}}]{wolfp2003}
{Wolf}, S., {Padgett}, D.~L., \& {Stapelfeldt}, K.~R. 2003, \apj, 588, 373

\bibitem[{{Youdin} \& {Goodman}(2005)}]{youdin05a}
{Youdin}, A.~N. \& {Goodman}, J. 2005, \apj, 620, 459

\bibitem[{{Youdin} \& {Shu}(2002)}]{youdin02a}
{Youdin}, A.~N. \& {Shu}, F.~H. 2002, \apj, 580, 494

\bibitem[{{Zhang} {et~al.}(2016){Zhang}, {Bergin}, {Blake}, {Cleeves},
  {Hogerheijde}, {Salinas}, \& {Schwarz}}]{zhang16a}
{Zhang}, K., {Bergin}, E.~A., {Blake}, G.~A., {et~al.} 2016, \apjl, 818, L16

\bibitem[{{Zhang} {et~al.}(2015){Zhang}, {Blake}, \& {Bergin}}]{zhang2015}
{Zhang}, K., {Blake}, G.~A., \& {Bergin}, E.~A. 2015, \apjl, 806, L7

\end{thebibliography}

\end{document}